\documentclass[aps,
groupedaddress,showpacs,nofootinbib,prl,twocolumn]{revtex4}

\usepackage{amssymb,amsmath}
\usepackage{graphicx}
\usepackage[english]{babel}
\usepackage{graphicx}
\usepackage{dcolumn}
\def\comment#1{}

\def\ns{\hspace{-1pt}}
\usepackage{bm}

\comment
{\usepackage{showlabels,rotating}

}
\input gbf.sty

\def\nablab{\mbox{\boldmath$\nabla$}}

\def\nablabf{\mbox{\footnotesize\boldmath$\nabla$}}

\def\dbar{~{\raisebox{.32em}{-}\hspace{-.57em}d}}

\def\N{{\cal N}}
\def\smbox#1{\hbox{\scriptsize#1}}
\newcommand{\sbar}{~\bar{}\!\!S}

\def\iems#1{}

\def\cl{{\rm cl}}

\def\fsz{\tiny}

\def\aut#1{#1}

\def\ins#1{{\it #1}}

\def\meq#1{}
\def\meq#1{}
\def\Tr{{\rm Tr}}
\def\iem#1{{\it #1}}

\def\ins#1{}

\newcommand{\lfrac}[2]{#1/#2}

\def\dst{\displaystyle}
\def\mn#1{\marginpar[]{\scriptsize#1}}
\def\mn#1{}

\def\IncludeEpsImg#1#2#3#4{\renewcommand{\epsfsize}[2]{#3##1}{\epsfbox{#4}}}

\usepackage{ulem}
\usepackage{cancel}
\usepackage{color}
\def\rr#1{\textcolor{red}{#1}}

\def\mn#1{\marginpar[\tiny{\rr{#1}}]{\tiny{\rr{#1}}}}

\def\rr#1{}

\usepackage{hyperref}

\def\IncludeEpsImg#1#2#3#4{\renewcommand{\epsfsize}[2]{#3##1}{\epsfbox{#4}}}

\newcommand{\be}{\begin{equation}}\newcommand{\ee}{\end{equation}}
\newcommand{\bea}{\begin{eqnarray}}\newcommand{\eea}{\end{eqnarray}}
\newcommand{\beaa}{\begin{eqnarray}}\newcommand{\eeaa}{\end{eqnarray}}
\newcommand{\ba}{\begin{array}}\newcommand{\ea}{\end{array}}
\newcommand{\bit}{\begin{itemize}}\newcommand{\eit}{\end{itemize}}
\newcommand{\ben}{\begin{enumerate}}\newcommand{\een}{\end{enumerate}}

 \newcommand{\sfrac}[2]{\raisebox{0.095ex}{\scriptsize${\frac{#1}{#2}}$}}
 \newcommand{\ssbf}[1]{\mbox{\tiny\bf{#1}}}
 \newcommand{\sbf}[1]{\mbox{\scriptsize\bf{#1}}}

\def\lfrac#1#2{#1/#2}

									\def\be{\begin{equation}}
\def\ee{\end{equation}}				\def\bea{\begin{eqnarray}}				\def\eea{\end{eqnarray}}
\def\bear{\begin{array}}				\def\eear{\end{array}}

												\def\x5{x^{5}}

\begin{document}

\title{
Effective Action and Field Equation for BEC from Weak to Strong Couplings}

\comment{
\author{H. Ruby}
\email{hruby@live.de}
\affiliation{ICRANeT Piazzale della Repubblica, 10 -65122, Pescara, Italy}
}

\author{Hagen Kleinert}
\email{h.k@fu-berlin.de}


\affiliation{Institut f{\"u}r Theoretische Physik, Freie Universit\"at Berlin, 14195 Berlin, Germany}
\affiliation{ICRANeT Piazzale della Repubblica, 10 -65122, Pescara, Italy}


\vspace{2mm}
\def\sch{Schr\"odinger}
\def\comment#1{}

\begin{abstract}

While free and weakly interacting nonrelativistic particles
are described by a Gross-Pitaevskii equation,
which is a
nonlinear self-interacting
Schr\"odinger equation,  the phenomena
in the strong-coupling limit
are governed
by an effective
action that is extremized by
a
double-fractional generalization of this equation.
Its particle orbits
perform L\'evy walks
rather than Gaussian
random walks.

\end{abstract}


\maketitle

  \section{Introduction}
  A quarter of a century after Bose and Einstein's 1924-1925 prediction \cite{BO,EI} that a free gas of Bose particles 
would condense at low temperature to a coherent state, 
Bogoliubov developed a first theory of weakly interacting
Bose-Einstein condensates (BEC) in
1947 \cite{BOG}.
It was an inspiring paper for many similar condensation phenomena,
such as superfluidity and superconductivity.
The experimental study of a BEC 
had to wait until 1995. Only then were cooling
and trapping  techniques sufficiently well developed
to prepare large enough samples 
of such gases to observe their physical properties. Presently these supply us with important
systems
on which we can test the theoretical tools developed in many-body physics 
and quantum field theory.

As an important result it was found  that
a nonlinear Schr\"odinger equation, which in this context is known as Gross-Pitaevskii equation,
yields satisfactory descriptions of  such condensed states
if the temperature is sufficiently low \cite{TEXT,GRP}. 
This is possible due to the high dilution of the condensates
which ensures the weakness of the interaction
of the particles.

More recently, however, 
it has been possible to increase the interaction strength
of these gases so much that in spite of the high dilution,
the interactions can be considered as strong.
The way to do is to exploit the so-called Feshbach resonance
in the two-body potential.
There exists a magnetic field $B_c$ where the scattering length 
of the  two-body potential
grows to infinity \cite{ZW}.  
In that regime, the Bogoliubov theory 
is no longer applicable,
and the quantum field theory of the Bose gas becomes completely nontrivial
and therefore interesting.

If we want to carry quantum field theoretical calculations
from the well understood weak-coupling regime to 
strong couplings, we may resort to various procedures.
One is, of course, the numerical simulation of the partition function
of the  system on a computer.
This may yield numerical date with can be compared with experiments.
A more satisfactory understanding can, however, be reached
by analytic methods.
If the action of the systems is formulated on a lattice,
there are high-temperature and low-temperature expansions.
The critical regime between them remains, however,
hard to reach since both expansions diverge where they are supposed to
meet.
The problem is that 
one needs in principle infinitely many orders to reach a critical
point.

In the past twenty years, methods have been developed to use weak-coupling 
expansions and resumming them in the strong-coupling regime.
This is known as the renormalization  group approach \cite{RG}.
If the field theory is formulated  in the continuum, the 
weak-coupling expansions require the calculation
of Feynman diagrams.
The critical point lies at an infinite coherence length
which corresponds to a zero mass of the euclidean field theory.
At that point, many 
Feynman diagrams diverge.
If we want to study the limiting theory, we must renormalize 
the theory. For this we introduce an arbitrary mass scale $\mu$, and for each 
 $\mu$ we calculate a renormalized coupling strength $g$.
The resulting expansions in powers of  $g$
are all divergent with a vanishing radius of convergence.
They need sophisticated mathematical methods of resummation.

There exists a
a {\it renormalization group function\/}  $\beta(g)$
which controls the change of the renormalized coupling constant under a change of the renormalization scale $\mu$ \cite{STP,GML}.
If this function
has a fixed point $g^*$ in the infrared limit $\mu\rightarrow 0$, then the resummed 
series yield reasonable
estimates for
the critical
behavior of a theory.
The general form of the physical laws in this limit have been formulated by many authors
\cite{WIDOM,STA,KAD}.  
 
A much  simpler and yet most  powerful method 
has been developed in the textbook \cite{KS22}. Again,
the basic input
consists in the diagrammatic expansions 
of the amplitudes,
but here expressed directly
in powers of the {\it un}\phantom{\hspace{1.pt}}renormalized or {\it bare}
coupling constant $g_B$.
These expansions are subjected to a 
variational
procedure \cite{FK} that converts the divergent weak-coupling power
series into convergent strong-coupling power series \cite{WS}.
Their expansion parameter is a certain inverse power of the coupling constant,
$1/g_B^\omega$. The parameter $\omega$ is the so-called exponent governing the 
approach to scaling, introduced by Wegner in 1990 \cite{WEGNER}.
The  new method is called {\it Variational Perturbation Theory} (VPT).
If applied to quantum mechanics it reproduces the same results as the 
old-fashioned
$\delta$-expansion \cite{DELTA}.
In quantum field theory, however, 
there are anomalous power laws, which cannot be handled by
 the $\delta$-expansion, and
the full 
 VPT is essential. It has so
far given the most accurate 
critical exponents measured by precision experiments 
in
micro-gravity satellite environment
\cite{LIPA}.

For various physical quantities and critical exponents,
the critical expressions and the approach to the critical
limit have been given in the textbook
\cite{KS22}.
It is the purpose of this paper to extend the method to the calculation of 
effective actions.
From perturbation theory it is well known how to calculate 
the effective action as a power series in the field strength, to be called 
$\Phi$.
This expansion is valid as long as $\Phi$
remains small. For large field strength, however, the effective action 
exhibits a nontrivial power behavior
which is approached in a way controlled again by the Wegner exponent $\omega$.
We shall present a method for finding a global expression
the effective action
as a function of 
 of any
 $\Phi$. It has  for small $\Phi$ the correct power series,
 and for large $\Phi$ the correct power behavior,
with the proper approach to the scaling limit.


  \section{Generating Functional and Effective Action}
The physical properties of a system described by a scalar quantum field $\phi(x)$
can be derived from a classical effective action that is derived as follows.
One starts from the generating functional of all Green functions
\begin{eqnarray} \label{d12.90}
   Z[j] = e^{iW [j]/\hbar },
\end{eqnarray}
where $W[j]$ is the
generating functional of all {\it connected\/} Green functions.
The vacuum expectation of the field, the average
\begin{equation} \label{d12.91a}
\Phi (x)\equiv \langle \phi (x)\rangle ,
\end{equation}
is given
by the first functional derivative
\begin{equation} \label{d12.91b}
\Phi (x)=   \delta W [j] /\delta j(x).
\end{equation}
This can be inverted to yield $j(x)$ as an $x$-dependent
 functional of $\Phi (x)$:
\begin{equation} \label{d12.j}
j(x)=j[\Phi ](x).
\end{equation}
This is used to form the Legendre transform of $W[j]$:
\begin{equation} \label{d12.G}
\Gamma [\Phi ]\equiv W[j]-\int d^4xj(x)\Phi (x) .
\end{equation}
where $j(x)$ on the right-hand side is replaced by (\ref{d12.j}).
This is the 
{\it effective action\/}
\iems{effective action}%
of the theory.
Its first functional derivative gives back the current
\begin{eqnarray} \label{d12.92}
\frac{
\delta \Gamma [\Phi ] }{\delta \Phi (x)} = -j(x).
\end{eqnarray}
In the absence of an external current $j(x)$, the effective action is extremal
on the physical field expectations $\Phi(x)$.

The generating functional of all
connected
 Green functions can
be recovered from the effective action
by the inverse Legendre transform
\begin{eqnarray} \label{d12.90a}
  W [j]  =  \Gamma [\Phi ]
             + \int d^4x j(x) \Phi (x).
\end{eqnarray}

In general, the above quantities
can be obtained 
with the help of
functional integrals.
The 
generating functional $Z[j]$, for example,  is given by
\begin{eqnarray} \label{d12.93a}
Z[j]
        = \frac{ \int {\cal D} \phi (x) e^{(i/\hbar ) \left\{ {\cal A}
          [\phi ]+ \int d^4x j(x)\phi(x) \right\} }}
          {\int {\cal D} \phi (x) e^{(i/\hbar ){\cal A}_0 [\phi ]}}.
\end{eqnarray}
Using (\ref{d12.90})
and 
 (\ref{d12.G}), this amounts to the functional integral formula
for $ \Gamma[\Phi]$:
\begin{eqnarray} \label{d12.93} \label{@9}
&&  \!\!\!\!\!\!\!\!\!\!\!\!\! e^{\frac{i}{\hbar} \left\{ \Gamma [\Phi ]
        + \int d^4x j(x)\Phi(x) \right\} }\nonumber \\
  &&      =\N{ \int {\cal D} \phi (x) e^{(i/\hbar ) \left\{ {\cal A}
          [\phi ]+ \int d^4x j(x)\phi(x) \right\} }}
\end{eqnarray}
with the
normalization factor
\begin{eqnarray} \label{d12.93b}
{\cal N} =
     \left\{ \int {\cal D} \phi (x) e^{(i/\hbar )
      {\cal A}_0 [\phi ]}\right\}^{-1}.
\end{eqnarray}
In writing this we have explicitly displayed
the  fundamental action quantum  $\hbar$, which is
 a measure for the size of quantum
fluctuations. There most physical systems,
the quantum fluctuations are rather small except in the
immediate vicinity of critical points. 
\comment{For these systems,
it is desirable to develop a method of evaluating (\ref{d12.93})
as an expansion in powers of $\hbar$,
which amounts to taking into account successively increasing
quantum fluctuations.
}

For $\hbar \rightarrow 0$, the path integral over the field $\phi (x)$
in
(\ref{d12.93a})
is dominated by the classical solution $\phi _{\cl}(x)$
which extremizes the exponent
\begin{eqnarray} \label{d12.94}
\left .   \frac{\delta {\cal A}}{\delta \phi }\right \vert_{\phi =
           \phi _{\cl}(x)} = -j(x).
\end{eqnarray}
At this level we therefore identify
\begin{eqnarray} \label{d12.98a}
W[j]&=&   \Gamma [\Phi ] + \int d^4x j(x)\Phi(x)\nonumber \\
& = & {\cal A} [\phi _{\cl}]
        + \int d^4x j(x)\phi _{\cl} (x).
\end{eqnarray}
Of course, $\phi _{\cl}(x)$ is a functional of $j(x)$,
so that we may write it more explicitly, as $\phi _{\cl} [j](x)$.
By differentiating  $W[j]$ with respect to $j$,  we have
from the general first part of Eq.~(\ref{d12.91a}):
\begin{eqnarray} \label{d12.99}
   \Phi (x) = \frac{ \delta W}{ \delta j } =
              \frac{\delta \Gamma }{\delta \Phi }
         \frac{\delta \Phi }{\delta j} + \Phi + j \frac{\delta \Phi }{
         \delta j} .
\end{eqnarray}
 Inserting the classical field equation (\ref{d12.94}),
this becomes
\begin{eqnarray} \label{d12.100}
    \Phi (x) =
   \frac{\delta {\cal A}}{\delta \phi _{\cl}} \frac{\delta \phi _{\cl}}
        {\delta j} + \phi _{\cl} + j \frac{\delta \phi _{\cl}}
        {\delta j} = \phi _{\cl}.
\end{eqnarray}
Thus, to this approximation,
   $\Phi(x) $ coincides  with the classical field
$\phi  _{\cl}(x)$. Replacing $\phi  _{\cl}(x)\rightarrow\Phi(x) $
 on the right-hand side of Eq.~(\ref{d12.98a}), we therefore obtain the
 lowest-order  result (zeroth order in $\hbar $):
\begin{eqnarray} \label{d12.101} \label{@15}
  \Gamma [\Phi ] = {\cal A} [\Phi ]
\end{eqnarray}
i.e., the effective action equals the fundamental
action.

In general it can be shown that
the effective action can be expanded in a functional power series of $\Phi(x)$
\begin{eqnarray}
\Gamma[\Phi]\!\!\!
&=&\!\!\!\sum_{n=0}^\infty\frac1{n!}\int d^Dx_1d^Dx_2\cdots
d^Dx_n
\Gamma^{(n)}(x_1,x_2,\dots,x_n)
\nonumber \\
&\times&\Phi(x_1)\Phi(x_2)\cdots
\Phi(x_n).
\label{@}\end{eqnarray}
where the $n$th coefficient
is just the
$n$-point vertex function 
of the theory  \cite{PI,KS22}.

For the $\phi ^4$-theory with O({$N$)}-symmetry and action
\begin{eqnarray} \label{d12.102B}
\!\!\!\!\!\!\!\!
 {\cal A} [\phi] =\int d^4x \left[ \frac{1}{2} \left( \partial \phi _a\right) ^2
             - \frac{m^2}{2} \phi _a^2 - \frac{g}{4!}
             \left( \phi _a^2\right) ^2\right],
\end{eqnarray}
the zeroth-order effective action is
\begin{eqnarray} \label{d12.102}
\!\!\!\!\!\!\!\!
  \Gamma  [\Phi] =\int d^4x \left[ \frac{1}{2} \left( \partial \Phi _a\right) ^2
             - \frac{m^2}{2} \Phi _a^2 - \frac{g}{4!}
             \left( \Phi _a^2\right) ^2\right].
\end{eqnarray}
For $m^2>0$,%
\footnote{%
The case $m^2 <0$ corresponds to the condensed phase of the system
and will be treated below.}
this has
an extremum at $\Phi_a \equiv { 0}$, and
there are only
two non-vanishing vertex functions
$ \Gamma ^{(n)} (x_1, \dots , x_n) $:
\\
\textbf{\textit{n}}=2:
\begin{eqnarray} \label{d12.103}
\!\!\!\!\!\!\!\!\!\!\!\!\!\!
 \Gamma ^{(2)}_{ab} (x_1, x_2) &\equiv &\left.
\frac{\delta ^2\Gamma }{\delta \Phi _a
(x_1) \delta \Phi _b (x_2) }\right\vert_{\Phi _a = 0}\nonumber \\
&=&
\left.\frac{\delta ^2{\cal A}}{ \phi_a (x_1) \phi_b (x_2)  }
      \right\vert_{\phi _a = \Phi _a =0}
\nonumber \\
&          = &(- \partial ^2 -m^2) \delta _{ab}
	 \delta ^{(4)} (x_1 - x_2) . \label{@17}
\end{eqnarray}
This determines
the inverse of the propagator:
\begin{eqnarray} \label{d12.103x}
 \Gamma ^{(2)} _{ab}(x_1, x_2)=
       [i\hbar G^{-1}]_{ab}(x_1,x_2),
\end{eqnarray}
Thus we find to this zeroth-order approximation
that $G_{ab}(x_1,x_2)$ is equal to the free propagator:
\begin{equation}
G_{ab}(x_1,x_2)=
G_{0{ab}}(x_1,x_2)
\label{@}\end{equation}
~\\
\\
\textbf{\textit{n}}=4:
\begin{eqnarray} \label{d12.104}
   \Gamma ^{(4)}_{abcd}(x_1, x_2, x_3, x_4)
 \!\!\!&  \equiv&\!\!\! \frac{\delta ^4\Gamma }{\delta \Phi _a (x_1\hspace{-1pt})
        \delta \Phi _b (x_2\hspace{-1pt}) \delta \Phi _c (x_3\hspace{-1pt})\delta \Phi _d (x_4\hspace{-1pt})}\nonumber \\& =& g
        T_{abcd}  ,
 \label{@20}\end{eqnarray}
with
\begin{equation} \label{d12.181b}
\!\!\!\!\!\!\!\!\!\!\!\!
T_{abcd}=\frac{1}{3}(\delta _{ab}\delta _{cd}+\delta _{ac}\delta _{bd}+
\delta _{ad}\delta _{bc}),
\end{equation}
which is just the fundamental vertex
implied by the local interaction (\ref{d12.102}).

According to the definition of the effective action,
all diagrams of the theory can be composed from the propagator
$G_{ab}(x_1,x_2)$ and this vertex
via tree diagrams. Thus we see that in this lowest approximation,
we recover precisely the subset of all original Feynman
diagrams with a tree-like topology. These are all diagrams which do not
involve any loop integration.
Since the limit $\hbar \rightarrow 0$
    corresponds to the classical equations of motion with no
    quantum fluctuations we conclude: Classical
    field theory corresponds to quantum field theory 
in the tree approximation.

The use of the initial action as an approximation to the
 effective action  neglecting
 fluctuations is often referred to as {\it mean-field theory\/}.

Let us now apply the effective action formalism
to the phenomenon of Bose-Einstein 
condensation (BEC). There the number of O($N$) components is $N=2$,
and $\Phi^2$ is identified with $\Psi^\dagger\Psi/Mk_BT$,
and the interaction with $(g_{S}/2)\int d^D(\Psi^\dagger\Psi)^2$, the subscript $S$ indicating that the lowest-order
equation of motion of the field $\Phi$ becomes the nonlinear Schr\"odinger equation
\begin{eqnarray}
\left[-i\hbar\partial_t-\frac{\hbar ^2}{2M}\nablab^2-\mu+g_S\Psi^\dagger\Psi\right]\Psi(x)=0,
\label{@GPE}
\label{@22}\end{eqnarray}
where $M$ is the physical mass of the particles.
The relation with the previous coupling constant is
\begin{eqnarray}
g_S/2=g/4!.
\label{@gSg}\end{eqnarray}
Here $\mu$ is the chemical potential of the particles
which is fixed by ensuring a given particle number $N$.
 
After quantization,  the nonlinear Schr\"odinger equation
(\ref{@22}) has a phase transition at a temperature $T_c$
where the  chemical potential  $\mu$ vanishes.
For a free gas,
this is determined by the equation
\begin{eqnarray}
\!\!N=
\sum _{\sbf p}
\frac{1}{e^{  \beta \hbar  \omega _{\ssbf p} - \beta \mu }-1},
\label{@AVNUb}
\end{eqnarray}
where $\beta=1/k_BT$, $k_B$ is the Boltzmann constant, and
$\omega _{\sbf p}={\bf p}^2/2M$ the kinetic energy of the particles. In $D$ dimensions, (\ref{@AVNUb}) can be written
as 
an energy integral
\begin{eqnarray}
\!\!N
&=&\frac{1}{ \Gamma (D/2)}
\frac{V_D}{\sqrt{2\pi  \hbar ^2 \beta /M}^{D}}
  \int_{0}^{\infty} d \varepsilon \frac{ \varepsilon ^{D/2-1}}
{e^{ \varepsilon - \beta \mu }-1}
\comment{
\nonumber \\
&=&\frac{1}{ \Gamma (D/2)}
\frac{V_D}{\sqrt{2\pi  \hbar ^2 \beta /M}^{D}}\zeta(D/2),
}
\label{@AVNUb2}
\end{eqnarray}
where $V_D=$ volume.
For $\mu=0$, the integral
yields Riemann's zeta function
 $\zeta(D/2)=\sum_ {n=1}^\infty n^{-D/2}$, and we 
find the critical temperature from
\begin{eqnarray}
\!\!N
&=&\frac{1}{ \Gamma (D/2)}
\frac{V_D}{\sqrt{2\pi  \hbar ^2 \beta_c /M}^{D}}\zeta(D/2),
\label{@AVNUb3}
\end{eqnarray}
In the neighborhood of $T_c$ and for $D=3$, the chemical potential behaves like
\begin{equation} \!\!\!\!\!\!\!\!
-\mu  \approx
\frac{1}{4\pi} k_BT_c\,
{ \zeta^2 (3/2)}
{\left[\left(\frac{T}{T_c}\right)^{3/2}-1\right] }^2.
\end{equation}
This is the  {\it critical regime}
of the Bose gas.
Its statistical properties are governed by the action
\begin{eqnarray}
{\cal A}\!=\!\beta\!\int d^Dx \left[-\frac{\hbar ^2}{2M}\partial\Psi^*\partial\Psi+ \mu\Psi^*\Psi-\frac{g_S}2(\Psi^*\Psi)^2 \right]\!.
\label{@29}\end{eqnarray}
The coupling constant $g_S$ can be 
expressed in terms of the $s$-wave scattering length $a_s$
of the atoms as
$Mg_S=4\pi \hbar ^2 a_s$
\cite{GRP}. The connection with 
the action (\ref{d12.102B}) is established 
by equating $m^2=-2M\mu$ and $g=4!\, 2\pi a_s M k_BT$ which has 
near $T_c$ is, in natural units with $\hbar=1$, $c=1$, the approximate size
$g\approx 
4!\,2\pi a_s/\ell_c^2$
where $\ell_c=[\zeta(D/2)V_D/N \Gamma(D/2)]^{1/D}$.

If the BEC takes place in an external trap
potential $V({\bf x})=M \omega^2 {\bf x}^2/2$,
the action (\ref{@29})
contains, instead of the chemical potential $\mu$,
an $ {\bf x}$-dependent chemical potential
 $\mu( {\bf x})=\mu+M \omega^2 {\bf x}^2/2$, and the particle density 
$n( {\bf x})=\Psi^*\Psi$ is obtained
from the solution of the effective field equation
\begin{eqnarray}
\left[
-\frac{\hbar ^2}{2M}\nablab^2+\mu+M \omega^2 {\bf x}^2/2+g_Sn({\bf x})\right]\Psi(x)=0.
\end{eqnarray}
In the Thomas-Fermi approximation,
we neglect the gradient term and find the density profile
\begin{eqnarray}
n({\bf x})=\frac{M\omega^2}{g_S}(R_c^2-R^2),~~~R^2= {\bf x}^2,
\label{@}\end{eqnarray}
where ${M\omega^2}R_c^2\equiv-\mu$.

The interaction shifts the critical temperature of the BEC
in $D=3$ dimensions
from the free-particle value determined by (\ref{@AVNUb3})
to a higher value $T_c+\Delta T_c$
with $\Delta T_c/ T_c\approx a_s (N/V_3)^{1/3}$ \cite{HKFL}.

The reason for the good agreement with observations
is the smallness  of the coupling constant 
$g_S$ which, as we discussed before, 
is ensured in experimental BEC
by the diluteness 
of the atomic gases that can be brought to the low critical 
temperatures in the laboratory.
If the coupling gets larger, the interaction changes the 
properties of the gas, 
and the extremization of the action 
(\ref{@29})
will no longer be sufficient to explain the physical properties 
of the gas.
Instead, one has to extremize the effective action $\Gamma[\Phi]$,
to which (\ref{@29}) is merely the mean-field approximation.

An important experimental method to investigate
the strong-coupling properties 
of the gas is based on
 placing the sample in a magnetic field $B$
and raising this up to a critical value $B_c$, where the two-particle 
scattering length diverges at a so-called {\it Feshbach resonance.}
At that point one reaches the strong-coupling limit of the 
the field theory (\ref{@29}).
The purpose of this note is to calculate the 
full
effective action $\Gamma[\Phi]$ of the theory for all couplings,
from weak to strong.

\comment
{
If the interactions become stronger
The nonlinear
Schr\"odinger equation
 (\ref{@22}) is called the Gross-Pitaevskii equation.
It extremizes the 
action in the mean-field approximation. For 
small couplings $g$, thereby explaining most observed properties 
of the BEC. 
In the following we shall focus our attention upon time-independent systems 
in which case we can omit the first term of (\ref{@22}).
}

  \section{Quadratic Fluctuations}
Let us calculate  the higher  $\hbar$-correction to the mean-field approximation
(\ref{@15}). For this we
 expand
the action in (\ref{@9}) in powers of the fluctuations of the field around the
classical solution
\begin{eqnarray} \label{d12.95}
  \delta \phi (x)\equiv \phi (x) - \phi _{\cl} (x),
\end{eqnarray}
and perform a perturbation expansion.
  The quadratic term in $\delta \phi (x)$
  is considered as a ``free-field action'',
 the higher powers in $ \delta  \phi (x)$
as ``interactions''.
%
  %
  %
Up to second order in the fluctuations $\delta \phi (x)$, the
action is expanded as follows:
\begin{eqnarray}
&&
  \hspace{-3mm}
{\cal A} [\phi_{\cl}+  \delta \phi ] +\int d^4x j(x)
[\phi_\cl(x)+ \delta\phi(x)]\nonumber \\
&&=
  {\cal A} [ \phi _{\cl}]   +\int d^4x j(x)\phi_\cl(x)
\nonumber \\&& ~~~~~ \!~~~~~~~+ \left. \int
           d^4x \left\{j(x)+\frac{\delta {\cal A}}
                {\delta \phi (x)}
           \right\vert_{\phi =\phi _\cl}  \right\} \delta \phi (x)
\nonumber \\
&&
 ~~~~~~~\! ~~~~~ +  \int
           d^4x d^4y\left .\delta  \phi (x) \frac{\delta ^2{\cal A}}
                 {\delta \phi (x) \delta \phi (y)}
\right                  \vert_{\phi =\phi _\cl} \delta \phi (y)
            \nonumber \\&& ~~~~~ ~~~~~ ~~~~~     + {\cal O} \left.\left( (\delta  \phi )^3\right)\right. .
 \label{d12.96}
\end{eqnarray}
The curly bracket term that is linear in 
the variation $ \delta \phi$ vanishes
 due to the
 extremality
 property of the classical field $\phi _{\cl}$
expressed by
the field equation  (\ref{d12.94}). Inserting this expansion into (\ref{d12.93}),
we obtain the approximate expression
\begin{eqnarray}
\label{fluctint3X}\!\!\!\!\!\!
&&\!\!\!\!\!\!\!\!\!\!\!\!Z[j] \approx \N
e^{(\lfrac{i}{\hbar})\left\{ {\cal A}[\phi _{\cl}] + \int
          d^4xj \phi _{\cl}\right\} } \int {\cal D}  \delta \phi          \\&\!\!\!\!\!\!&\!\!\!\!\!\!\times	 \exp\left\{\frac{i}{\hbar }
 \int
           d^4x d^4y\left .\delta  \phi (x) \frac{\delta ^2{\cal A}}
                 {\delta \phi (x) \delta \phi (y)}
\right                  \vert_{\phi =\phi _\cl} \delta \phi (y)
\right\}
\nonumber \end{eqnarray}

 We now observe that the fluctuations in $\delta \phi $
 will be of average size $\sqrt{ \hbar}$ due to the
 $\hbar$-denominator in the Fresnel integrals  over
 $ \delta \phi $ in (\ref{fluctint3X}). Thus
 the fluctuations $(\delta \phi )^n$ are on the average of relative
 order $\hbar^{n/2}$. If we ignore corrections of order
 $\hbar^{3/2}$, the fluctuations remain quadratic in $\delta \phi $
 and we may  calculate the right-hand
 side of (\ref{fluctint3X}) as
\begin{eqnarray} \label{d12.97}
  && \!\!\!\!\!\!\!\!\!\!\!\!\!\!\!\!\nonumber 
\N
e^{(\lfrac{i}{\hbar})\left\{ {\cal A}[\phi _{\cl}] + \int
          d^4xj(x) \phi _{\cl}(x)\right\} } \left[\det\frac{ \delta ^2 {\cal A} }{\delta \phi
          (x) \delta \phi (y)}\right]_{\phi =\phi _{\cl}}
 \\
   &&=\left( \det iG_0\right) ^{1/2}  e^{ (\lfrac{i}{\hbar} )\left\{ {\cal A} [\phi _\cl]
            + \int d^4x j(x)\phi _{\cl}(x) 
 \right\}
}  \nonumber\\
&&\times~
 e^{ (\lfrac{i}{\hbar} )\left\{  i (\lfrac{\hbar}{2})
              \smbox{Tr} \log [\delta ^2 {\cal A} / \delta \phi (x) \delta
              \phi
                 (y) \vert_{\phi =\phi _{\cl} } \right\} }.
\end{eqnarray}
Comparing this with
the left-hand side of
(\ref{d12.93}), we find that to first order in $\hbar$, the effective action may be
recovered by  equating
\begin{eqnarray} \label{d12.98}
\!\!\!\!\!\!\!\!\! \!\!\!\!\!\!\!\!\!\!
&&\! \!\!\!\!\!\!\!\!\!\!  \!\!\!\! \!\!\!\!  \Gamma [\Phi ] + \int d^4x j(x)\Phi(x) =  {\cal A} [\phi _{\cl}]
        + \int d^4x j\phi _{\cl}
\nonumber \\&&
         + \frac{i\hbar}{2} \Tr \log \frac{\delta ^2 {\cal A}}
               {\delta \phi (x) \delta \phi (y)}
               \left( \phi _{\cl}\right) .
\end{eqnarray}
In the
limit $\hbar \rightarrow0 $, the trace log term disappears
and (\ref{d12.98}) reduces to the classical action
as in (\ref{d12.101}).

To
 include the $\hbar$-correction into
$ \Gamma [\Phi]$,  we  expand $W[j]$ as
\begin{eqnarray} \label{d12.105}
 W [j] = W _0 [j] + \hbar W _1 [j] +{\cal O}( \hbar^2) .
\end{eqnarray}
Correspondingly, the field $\Phi $  differs from
 $\Phi_{\cl}$ by a correction of order $\hbar^2$.
\begin{eqnarray} \label{d12.106}
  \Phi = \phi _{\cl} + \hbar \phi _1 + {\cal O}
        (\hbar^2) .
\end{eqnarray}
Inserting this into (\ref{d12.98}), we find
\begin{eqnarray} \label{d12.107}~
 \Gamma [\Phi ] + \!\!\int d^4x\,j \Phi      \!\!& =&\!\!
   {\cal A} \left[ \Phi -\hbar \phi _1\right]
      \left.   + \int d^4xj \Phi -\hbar \int d^4xj \phi _1\right.
\nonumber \\
     & +& \left. \frac{i}{2} \hbar \Tr \log \frac{\delta ^2{\cal A}}
           {\delta \phi _a \delta\phi _b} \right \vert_{\phi =
            \Phi -\hbar \phi _1}
  + {\cal O} \left( \hbar^2\right) .
                        \nonumber
\end{eqnarray}
Expanding the action up to the same
 order in $\hbar$ gives
\begin{eqnarray} \label{d12.108}
  \Gamma [\Phi ] & = & {\cal A} [\Phi ] + \hbar
           \left\{ \frac{\delta{\cal A}[\Phi]}{\delta \Phi } - j\right\}
           \phi _1
           + \int d^4x\,j \Phi  \nonumber \\
    &&  + \frac{i}{2} \hbar \Tr \log
            \left .  \frac{\delta ^2 {\cal A}}{\delta\phi _a \delta \phi _b}
             \right    \vert_{\phi = \Phi }  + {\cal O} \left( \hbar^2\right).
      \end{eqnarray}
  But because of (\ref{d12.94}), the curly-bracket term is only
of order
   ${\cal O} ( \hbar^2) $, so that we find the one-loop
   form of the effective action
\begin{eqnarray}  \!\!\!\!\!\!  \!\!\!\!\!\!\!\!  \!\!\!\!\!\!\!\!  \!\!\!\!\Gamma [\Phi ]\!\! &=& \!\!\Gamma_0[\Phi ] + \Gamma _1 [\Phi ] \nonumber \\ &=&\!\!
  \int d^4x \left[ \frac{1}{2} \left( \partial \Phi_a \right) ^2
        - \frac{m^2}{2} \Phi _a^2 - \frac{g}{4!}
          \left( \Phi _a^2\right) ^2\right]    \nonumber   \\{}
   &     +&\!\! \frac{i}{2} \hbar \Tr \log \left[ -\partial^2 -m^2
            - \frac{g}{6} \left( \delta _{ab} \Phi _c^2
              + 2 \Phi _a \Phi _b\right) \right]\! .
\label{d12.109}
\end{eqnarray}
\comment{where we have dropped the  infinite additive constant
$
 \frac{i}{2} \hbar \Tr \log \left[ -\partial^2 -m^2
          \right]
. $
}
 In the special case of a one-component real field, this becomes
\begin{eqnarray} \label{d12.110}
  && \!\!\!\!\!\!\!\!  \!\!\!\!\! \! \!\!\!\!\!\!\!\!\Gamma [\Phi ]  =  \int d^4x \left[\frac{1}{2} (\partial \Phi )^2
        - \frac{m^2}{2} \Phi ^2 - \frac{g}{4!} \Phi ^4\right]
  \nonumber   \\{}
   &     +& \frac{i}{2} \hbar \Tr \log \left[ -\partial^2 -m^2
            - \frac{g}{2}  \Phi ^2
              \right]\!
 .
\end{eqnarray}

What is the graphical content of the set of all Green
functions at this level? For $j=0$,
we find
that the minimum lies at
 $\Phi = \Phi^0  \equiv  \Phi _j =0$, as in the mean-field approximation.
  Around this minimum, we may expand the trace log in powers of $\Phi $,
  and obtain for $\Gamma^1[\Phi]$ the series:
\begin{eqnarray}\label{d12.111} \!\!\!\!\!\!&&  \!\!\!\!\!\! \!\!\!\!\!\!\frac{i}{2} \hbar \Tr \!\log
              \left(\! - \partial^2 \!-\!m^2\! -\!\frac{g}{2}
\Phi ^2 \right)\!\\&=& \frac{i}{2} \hbar \Tr \!\log \left( \!-\partial^2\! -\!m^2\right)
       \!  +\! \frac{ i}{2} \hbar \Tr \!\log
          \left( 1 \!+\! \frac{i}{\!-\partial^2 \!-\!m^2} ig \frac{\Phi ^2}{2}\right)\!.\nonumber 
\comment{            \\
      & =& i \frac{\hbar}{2} \Tr \!\log \left( -\partial^2 -m^2\right)
            - i \frac{\hbar}{2} \sum ^{\infty} _{n=1}
            \left( - i\frac{g}{2}\right) ^n \frac{1}{n}
             \Tr \left( \frac{i}{-\partial^2 -m^2} \Phi ^2\right) ^n.
\nonumber
}
 \end{eqnarray}
The second term can be expanded in powers of $\Phi^2$ as follows:
\begin{eqnarray}
            - i \frac{\hbar}{2} \sum ^{\infty} _{n=1}
            \left(\! - i\frac{g}{2}\right) ^n \!\frac{1}{n}
             \Tr \left( \frac{i}{-\partial^2 -m^2} \Phi ^2\right) ^n.
\nonumber\label{d12.111X}
 \end{eqnarray}
If we insert
\begin{eqnarray} \label{d12.113}
  G_0 = \frac{i}{-\partial^2 -m^2},
\end{eqnarray}
then 
 $\Gamma^1[\Phi]$
can be written as
\begin{eqnarray} \label{d12.114}\!\!\!
i\frac{\hbar}{2} \Tr \log \left(\! -\partial^2\! -m^2\right) \!-\!i \frac{\hbar}{2}
     \sum ^{\infty} _{n=1} \left( \!-i\frac{ g}{2}\right) ^n
      \frac{1}{n} \Tr \left( G_0 \Phi ^2\right) ^n \!.
\end{eqnarray}
More explicitly,
the terms with $n=1$ and $n=2$ read:
\begin{eqnarray}
 &&\!\!\!\!\!\! \!\!-\frac{\hbar}{2} g \int d^4x d^4y  \delta ^{(4)}(x\!-\!y)
              G_0 (x,y) \Phi ^2
        (y)  \\
  &&\!\!\!\!\!\!\!\!+ i\hbar \frac{g^2}{16} \int
        d^4x d^4y d^4 z  \delta ^4 (x\!-\!z)
         G_0 (x,y) \Phi ^2 (y) G_0 (y, z) \Phi ^2 (z).
 \nonumber
\label{d12.115}
\end{eqnarray}
The expansion terms of (\ref{d12.114})
for $n\geq 1$ correspond obviously to the Feynman diagrams\\[-.6cm]
\begin{eqnarray} \label{fluctint2}\label{@38}
~~~~~~~~~~~\raisebox{-.6cm}{\input loops1.tx }
\end{eqnarray}
Thus the series (\ref{d12.114})
is  a  sum of all diagrams formed from one loop
and any number of fundamental $\Phi ^4$-vertices.

This type of loop expansion has been used for many years in the quantum field theory
of many-particle systems where it is known as
{\it Belyaev expansion} \cite{BELYE}.

To systematize the entire expansion
(\ref{d12.114}),
the trace log term
may be
pictured by a trivial
single-loop diagram without an extra vertex:
\begin{eqnarray} \label{fluctint2s}
\!\!\!\!\!i\frac{\hbar}{2} \Tr \log \left( -\partial^2 -m^2\right) =
\input loop0.tx ~~.
\end{eqnarray}

The first two diagrams in (\ref{fluctint2}) contribute
corrections
 to the vertices $\Gamma ^{(2)}$ and $\Gamma ^{(4)}$
of (\ref{d12.103}), (\ref{d12.104}). The remaining ones
produce higher vertex functions and
lead to more involved tree diagrams. Note that only
the first two corrections are formally divergent,
 all following loop integrals converge.
 In momentum space
we find from (\ref{d12.115})
\begin{eqnarray}
   \Gamma ^{(2)}(q) &=& q^2 - m^2 -\hbar \frac{g}{2} \int
          \frac{dk^4}{(2\pi )^4} \frac{i}{k^2 - m^2 +i\eta }
    \label{d12.118b0}
   \\
&&  \hspace{-5em}  \Gamma ^{(4)} (q_i)  = g - i \frac{g^2}{2}\times
  \\&&   \label{d12.118b}\hspace{-5em}  
            \left[ \int \frac{d^4k}{(2\pi )^4}
            \frac{i}{k^2\!-\!m^2\!+\!i\epsilon }
            \frac{i}{\left( q_1\! + \!q_2 \!-\!k\right) ^2\! \!-\!\!m^2 \!+\! i\eta }\nonumber 
 \!+\! 2 \mbox{~perm}\right]\! .
   \label{d12.118b}
\end{eqnarray}
The convergence of all higher diagrams in the
expansion  (\ref{fluctint2})
is ensured by the renormalizability of the theory
since only up to $n=4$
does one have the possibility to add counter terms of the same
form as the original Lagrangian. We may write  (\ref{d12.118b0})
in euclidean form as
%
\begin{eqnarray} \label{d12.120}
  \Gamma ^{(2)} (q)
           & = & \left( q^2 + m^2 + \hbar \frac{g}{2}
                 D_1\right), \\
           \Gamma ^{(4)} (q_i) & = & g - \hbar
                   \frac{g^2}{2} \left[ I \left( q_1 + q_2\right)
                   + 2 \mbox{~perm}\right].
\end{eqnarray}
\newcommand{\db}{\,\,{\bar {}\!\!d}\!\,\hspace{0.5pt}}%
where $D_1$ and $I(q)$ are the Feynman integrals
\begin{eqnarray} \label{d12.120A}D_1 =\int \db^4k_E \frac{1}{k^2_E + m^2}
\end{eqnarray}
and  
\begin{eqnarray} \label{d12.120B}I (q)=\int \db^4 k_E \frac{1}{\left( k^2_E+m^2\right) }
                  \frac{1}{\left[ (k+q)_E^2 + m^2\right] }.
\end{eqnarray}
The integrals can be calculated in $D$ dimensions by separating 
them into a
directional and a size integral as
\begin{eqnarray}
\int \db^Dk= \int \frac{d^Dk}{(2\pi)^D}&=&\frac{S_D}{(2\pi)^D}\int dk k^{D-1}\nonumber \\
&\equiv&
 \frac{\sbar_D}2 \int dk^2 (k^2)^{D/2-1}.
\label{@}\end{eqnarray}
We further simplify all calculations by performing a Wick rotation
of all energy integrals 
$\int dk_0$ into  $i\int_{-\infty} ^\infty dk_4$.
Then the integrals $\int d^4 k$ become what are called euclidean integrals 
$i\int d^4 k_E$ where $k_E^\mu=({\bf k},k_4)$, and $k^2
=k_0^2-{\bf k}^2$ becomes $-k_E^2=-({\bf k}^2+k_4^2)$.

By the same token we introduce the euclidean version
 $\Gamma_E[\Phi]=-i\Gamma[\Phi] $
of the
effective action (\ref{d12.110}) 
whose functional derivatives are vertex functions 
$\Gamma^{(n)}$ by formulas like 
(\ref{@17}) and (\ref{@20}). 

We further introduce renormalized fields
\begin{eqnarray}
\phi_R(x)\equiv
Z^{-1/2}_\phi\phi(x)
\label{@55}\end{eqnarray}
where $Z^{1/2}_\phi$ is a field renormalization constant.
It serves to absorb infinities arising  in the momentum integrals.
The renormalized vertex functions 
are obtained by calculating all
 vertex functions
in $D=4-\epsilon$ dimensions 
and fixing the renormalization constants
order by order in perturbation theory.
Alternatively, we can add to the bare action suitable 
counter terms.
In either way, we arrive at
finite
expressions, such as
\comment{
\begin{eqnarray} \label{d12.121}
\Gamma  ^{(2)}_R (q)& = & ( q^2_E + m^2),  \\
  \Gamma _R^{(4)} (q_i) & = & g - 
\hbar \frac{g^2}{2}
          \left\{ \left[ I \left( q_1 + q_2\right)
          + 2 {\rm~perm}\right] - 3 I(0)\right\}. \label{d12.122}
\end{eqnarray}
If we
regularize the integrals
dimensionally
in $4- \epsilon $ dimensions,
 the
counterterms take the forms
calculated in
Section \ref{dimreg},
 and (\ref{d12.121}) becomes
}
\begin{eqnarray} \label{d12.123}\!\!\!\!\!\!\!\!\!\!
    && \hspace{-6em}\Gamma _R^{(2)} (q)  =  \left.\Bigg\{ q^2 + m^2 +
             \frac{ \hbar}{2}g\mu^{-\epsilon} \sbar_D
              m^2 \right .\nonumber \\
        &&\mbox{}\hspace{-5em} \left .    \times
\left[ \frac{1}{2} \Gamma (2-\epsilon /2)
        \Gamma
              \left( -1 + \epsilon /2 \right) \left( \frac{m^2}
              {\mu ^2}\right) ^{-\epsilon /2}+\frac{1}{\epsilon }
              \right] \right.\Bigg\}, \\
\hspace{-1em}       \Gamma ^{(4)}_R (q_i) & = & g - \hbar \frac{g^2}{2}
              \left[ I_R \left( q_1 + q_2\right) +
2 {\rm ~perm}\right].
              \label{d12.124}
\end{eqnarray}
 where 
 %
\begin{eqnarray} \label{d12.126}
 I_R (q) = -\frac{1}{2} \sbar_D \mu ^{-\epsilon }
            \left[ 1 + L^m (q) + \log \frac{q^2}{\mu ^2}\right]
            + {\cal O}(\epsilon ),
\end{eqnarray}
with 
\begin{eqnarray} \label{9.147}
  && \!\!\!\!\!\!\!\! L^m(q)  =  \int^{1}_{0} dx \log \left[ x (1-x)+
                 \frac{m^2}{q^2}\right] \\
   & &\!\!\! =   - 2 + \log \frac{m^2}{q^2} + \frac{\sqrt{ q^2 +4m^2}}
              {\sqrt{ q^2}} \log \frac{\sqrt{ q^2 +4m^2} +\sqrt{ q^2}}
               {\sqrt{ q^2 + 4m^2}-\sqrt{ q^2}}. \nonumber
    \end{eqnarray}
In any regularization scheme, we can also perform 
subtractions 
of counter terms which all have 
the same form as the terms in the original action 
to remove the divergent parts 
of the Feynman integrals. 
In this way we obtain
for the the euclidean version $\Gamma_E[\Phi]=-i\Gamma[\Phi] $
of the
effective action (\ref{d12.110}) 
in $D=4-\epsilon $ dimensions 
the finite 
subtracted expression
\begin{eqnarray}   \Gamma_E [\Phi ] & = & \int d^Dx_E \left\{ \frac{1}{2}
           (\partial \Phi )^2 + \frac{m^2}{2} \Phi ^2 + \frac{g}
{4!}
            \Phi ^4 \right\}
\nonumber \\
&-&
         \mbox{}  \frac{\hbar}{2} \Tr \log \left( -\partial^2 +m^2
             +\frac{g}{2} \Phi ^2\right) \nonumber \\
         &+& \mbox{}  \frac{\hbar g}{4} \int \frac{d^Dq_E}{(2\pi )^4}
              \frac{1}{q_E^2 + m^2} \int d^Dx \, \Phi ^2 (x)
\nonumber \\
&-&
            \frac{\hbar g^2}{16} \int \frac{d^Dq_E}{(2\pi )^4}
                 \frac{1}{\left( q_E^2 + m^2\right)^2 } \int d^D x \, \Phi ^4.
\label{d12.128}
\end{eqnarray}
In this formulation, the divergent integrals in the last two terms
modify the mass term and the coupling constant. They may be
evaluated with any regularization method.
A direct evaluation 
in $4- \epsilon$ dimensions yields
\begin{eqnarray}
 \!\!\!\!\!\!\!\!\!\!\!\!\!\!  \Gamma_E [\Phi ] & = & \int d^Dx_E 
\left\{ \frac{1}{2} (\partial \Phi )^2
           + \frac{m^2}{2} \Phi ^2 +\frac{g}{4!} \Phi ^4\right\}
  \nonumber \\&-&        \frac{\hbar}{2} \Tr \log \left( -\partial^2
           + m^2 + \frac{g}{2}\Phi ^2\right) \nonumber \\
        \!\!\!\!\!\!\!\!\!\!\!\!  ~~~~~~~~~~~~~~~  &-&{}^mC_1 \frac{m^2}{2}
              \Phi ^2 -{}^gC_2 \frac{g^2}{4!} \Phi ^4,
 \label{d12.127}\end{eqnarray}
where the third line may be written as
\begin{eqnarray} \label{d12.134sec}\!\!\!\!\!\!
\!\!\!          && \mbox{}\!\!\!\!\!\!\!\!\!\! \!\!\!\!\!\!\!\!\!\!\!\!\!- \frac{\hbar g}{4} ( m^2)^{1-\epsilon/2} c_1\Phi ^2 + \frac{\hbar g^2}{16}(m^2)^{-\epsilon/2}
          c_2 \Phi ^4,
\end{eqnarray}
with the constants
\begin{eqnarray}
c_1=m^{\epsilon-2}\int \db ^Dp_E\frac{1}{p^2_E+m^2},\label{@Inc} \\
c_2=m^{\epsilon}\int \db ^Dp_E\frac{1}{(p^2_E +m^2)^2}.
\label{@Inc'}\end{eqnarray}
We evaluate these integrals using the formulas \cite{KS22}
%
\begin{eqnarray} \label{d12.137}
\!\!\!\!\!
  \int\! \!\!\dbar^D k _E\frac{1}{k^2_E\! +\! m^2 }
 \!  =\! \sbar_D\hspace{-1pt} \frac{\Gamma (D/2) \Gamma (1\!-\!D/2)} {2\Gamma(1)}
      \frac{1}{\left( m^2\right) ^{1\!-\!D/2}},
\end{eqnarray}
and
\begin{eqnarray} \label{d12.137b}
\!\!\!\!\!
  \int\ns \!\!\hspace{-1pt}\dbar^D k_E \frac{1}{(k^2_E\! +\! m^2 )^2}
 \!\ns  =\! \sbar_D \hspace{-1pt} \frac{\Gamma (D/2) \Gamma (2\!-\ns\!D/2)} {2\Gamma(1)}
      \frac{1}{\left( m^2\right) ^{2\!-\!D/2}}\!.
\end{eqnarray}
Further, by integrating (\ref{d12.137}) over $m^2$, we find
\begin{eqnarray} \label{d12.138}
\!\!\!\!  \int\hspace{-1pt} \hspace{-1pt}\!\!\dbar^D k_E \!\log\!\ns\left( k^2_E \!\ns+\! m^2\right) \!\! =\! \sbar_D\hspace{-1pt}
         \frac{\Gamma (D/2)  \Gamma (1\!-\!D/2)}
              {D\Gamma (1)}\! \left( m^2\right)^{D/2}\!\!\! .
\end{eqnarray}
In the so-called minimal subtraction scheme
in $4- \epsilon$ dimensions, only the singular
$1/ \epsilon$ pole parts  of
the two integrals
are selected for the subtraction in (\ref{d12.128}).
In the neighborhood 
of $\epsilon=0$,
(\ref{@Inc}) and (\ref{@Inc'})
become
\begin{eqnarray}
c_1&=&
- \sbar_D\frac2{\epsilon}+{\cal O}(\epsilon), \label{d12.135a}\\
c_2&=& \sbar_D\frac1{\epsilon}\left(1-\frac{\epsilon}{2}\right)+{\cal O}(\epsilon). 
\label{@Incp'}\end{eqnarray}
Hence we can choose
the singular terms in (\ref{d12.134sec})
\comment{behaves like
\begin{eqnarray} \label{d12.139}
  \sbar_D \left[ -1/\epsilon + {\cal O} (\epsilon )
            \right] \frac{1}{2} \left( 1 +\frac{ \epsilon}{4} \right)
        \!    \left( m^2\right) ^{\frac{2-\epsilon }{2}}\!\!\!,
\end{eqnarray}
If we attempt to take this expression to the $m=0$\,-limit,
the counter terms are no longer defined.
 We now see an important advantage of the $\epsilon$-expansion:
 The}
 %
\begin{eqnarray} \label{d12.135}
  - \frac{\hbar g\mu ^{-\epsilon }}{4} m^2 \sbar_D
         \left( -\frac{1}{\epsilon }\right)  \Phi ^2,~~~
         \frac{\hbar g^{2} \mu ^{-\epsilon }}{16} \sbar_D
         \frac{1}{\epsilon } \Phi ^4.
\end{eqnarray}
as counter terms.
These make the effective action (\ref{d12.127})
finite
for any mass $m$.

 Note, however, that an auxiliary mass parameter  $\mu $
must be introduced to define these expressions. If the physical mass
 $m$ is  nonzero, $\mu $ can be chosen to be equal to $m$.
But for $m=0$, we must use an arbitrary nonzero auxiliary mass $\mu$
as
the
renormalization scale.

Observe that up to the order $\hbar$, there is no divergence 
that needs to be absorbed in the gradient term $\int d^Dx\,(\partial \Phi)^2$ 
of the effective action (\ref{d12.127}).
These come in as soon as we 
carry the same analysis to one more loop order.

  Let us calculate the effective potential
in the critical regime for a constant
  field $\Phi $ at the one-loop level.
 It is defined by
   $v(\Phi ) = \lfrac{-\Gamma_E [\Phi ]}{VT}$.
 The argument
  in the 
trace log term is now diagonal in momentum space
  and the 
calculation  reduces to a simple momentum integral. 
It follows directly from
 (\ref{d12.127}) and reads
\begin{eqnarray} \label{d12.134}
     v(\Phi ) \!\!\! & = &\!\!\! \frac{m^2}{2} \Phi ^2\! +\! \frac{g}{4!} \Phi ^4\!              + \!\frac{\hbar}{2}\! \int\! \frac{d^Dq_E}{(2\pi )^D}
              \log \left( 1 + \frac{g}{2} \frac{\Phi ^2}{q^2_E \! +\! m^2}\!
              \right) \nonumber \\
      && 
\comment{\mbox{}\!\!\!\!\!\!\!\!\! \!\!\!\!\!\!\!\!\!\!\!\!\!- \frac{\hbar g}{4} \int  \frac{d^Dq_E}{(2\pi )^D}
            \frac{1}{q^2_E \!+\! m^2_E} \Phi ^2 + \frac{\hbar g^2}{16}
            \int \frac{d^Dq_E}{(2\pi )^D}\frac{1}{
            \left( q^2\!+\!m^2_E\right) ^2} \Phi ^4.\nonumber 
\end{eqnarray}
%
The second line may be written as
\begin{eqnarray} \label{d12.134secl}\!\!\!\!\!\!
        && \mbox{}\!\!\!\!\!\!\!\!\!\! \!\!\!\!\!\!\!\!\!}\!\!\!\!\!\!\!\!\!- \frac{\hbar g}{4} ( m^2)^{1-\epsilon/2} c_1\Phi ^2 + \frac{\hbar g^2}{16}(m^2)^{-\epsilon/2}
          c_2 \Phi ^4.
\end{eqnarray}
\comment{
where
\begin{eqnarray}
c_1=m^{\epsilon-2}\int \db ^Dp\frac{1}{p^2+m^2},\label{@Incl} \\
c_2=m^{\epsilon}\int \db ^Dp\frac{1}{(p^2 +m^2)^2},
\label{@Inc'l}\end{eqnarray}
We evaluate these integrals using the formulas \cite{KS22}
%
\begin{eqnarray} \label{d12.137}
\!\!\!\!\!
  \int \!\!\dbar^D k_E \frac{1}{k^2\! +\! m^2 _E}
 \!  =\! \sbar_D \frac{\Gamma (D/2) \Gamma (1\!-\!D/2)} {2\Gamma(1)}
      \frac{1}{\left( m^2\right) ^{1\!-\!D/2}}\!,
\end{eqnarray}
and
\begin{eqnarray} \label{d12.137b}
\!\!\!\!\!
  \int \!\!\dbar^D k_E \frac{1}{(k^2\! +\! m^2_E )^2}
 \!  =\! \sbar_D  \frac{\Gamma (D/2) \Gamma (2\!-\!D/2)} {2\Gamma(1)}
      \frac{1}{\left( m^2\right) ^{2\!-\!D/2}}\!.
\end{eqnarray}
Further, by integrating (\ref{d12.137}) over $m^2$, we find
\begin{eqnarray} \label{d12.138}
\!\!\!\!  \int \!\!\dbar^D k_E \log\left( k^2_E \!\!+\! m^2_E\right) \! =\! \sbar_D
         \frac{\Gamma (D/2)  \Gamma (1\!-\!D/2)}
              {D\Gamma (1)}\! \left( m^2\right)^{D/2}\!\! .
\end{eqnarray}
  In $4-\epsilon $ dimensions, the integrals 
(\ref{@Inc}) and (\ref{@Inc'})
become
\begin{eqnarray}
c_1&=&
- \sbar_D\frac2{\epsilon}+{\cal O}(\epsilon), \label{d12.135a}\\
c_2&=& \sbar_D\frac1{\epsilon}\left(1-\frac{\epsilon}{2}\right)+{\cal O}(\epsilon), 
\label{@Incp'}\end{eqnarray}
and
yield for the terms in (\ref{d12.134sec})
\comment{behaves like
\begin{eqnarray} \label{d12.139}
  \sbar_D \left[ -1/\epsilon + {\cal O} (\epsilon )
            \right] \frac{1}{2} \left( 1 +\frac{ \epsilon}{4} \right)
        \!    \left( m^2\right) ^{\frac{2-\epsilon }{2}}\!\!\!,
\end{eqnarray}
If we attempt to take this expression to the $m=0$\,-limit,
the counter terms are no longer defined.
 We now see an important advantage of the $\epsilon$-expansion:
 The}
the singular contributions
at $ \epsilon=0$, which we can remove with the help of the following 
counter terms:
\begin{eqnarray} \label{d12.135}
  - \frac{\hbar g\mu ^{-\epsilon }}{4} m^2 \sbar_D
         \left( -\frac{1}{\epsilon }\right)  \Phi ^2,~~~
         \frac{\hbar g^{2} \mu ^{-\epsilon }}{16} \sbar_D
         \frac{1}{\epsilon } \Phi ^4.
\end{eqnarray}
These regularize the effective potential
for any mass.
 Note, however, that an auxiliary mass $\mu $
had to be introduced to define these expressions. In the case
$m \neq 0$, $\mu $ can be chosen equal to $m$.
But for $m=0$, we must use an arbitrary nonzero auxiliary mass scale $\mu$.
}

The expansions in powers of $\epsilon $
has an important property which
has the direct applications  in the
description of the strong-coupling limit, i.e., 
in critical phenomena. For small 
$\epsilon $, Eq.~(\ref{d12.123})  can be rewritten as
\begin{eqnarray} \label{d12.129}\label{@62}
  \Gamma _R^{(2)} = \left( q^2_E + m^2  +\frac{ \hbar}4
          { \mu^{-\epsilon}g} \sbar_D \log \frac{m^2_E }{\mu ^2}\right)
\end{eqnarray}
which is,
to the same order in $g$,  equal to
\begin{eqnarray} \label{d12.130}
\label{@63}
  \Gamma ^{(2)}_R (q) =  \left[ q^2_E  + \left( \frac{m^2}{\mu ^2}\right)
         ^{1+\frac{\hbar}4 \mu^{-\epsilon}g \sbar_D} \mu ^2\right] .
\end{eqnarray}
This means that the vertex function at $q=0$
has a mass term 
that depends on the mass $m$ of the $\phi$-field
via a power law:
\begin{eqnarray} \label{d12.131}
 \label{@Form52}
  \Gamma _R^{(2)} (0) =  \left( \frac{m ^2}{\mu ^2}\right) ^\gamma
           \mu ^2.
\end{eqnarray}
The power 
$\gamma$
depends in 
the coupling strength $g$ like
\begin{eqnarray} \label{d12.132}\label{@65}
  \gamma  = 1 + \frac{\hbar}{4} \mu^{-\epsilon}g \sbar_D.
\end{eqnarray}
The important point is that this power $ \gamma$ is measurable
as an experimental
quantity called the \iem{susceptibility}.
It is 
called the {\it critical exponent}
\ins{critical,exponent}%
\ins{exponent;critical}%
 of the
susceptibility.

If the effective action
is calculated to order $\hbar ^2$,
then
the gradient term in the effective action is modified and becomes
\begin{eqnarray}
\Gamma[\Phi]=\int d^dx\,\Phi_R (x)\Gamma_R(\hat q)\Phi_R (x),
\label{@}\end{eqnarray}
where 
$
\Phi_R (x)=Z_\phi^{-1/2} \Phi (x)$, and $Z_\phi
$
is the {\it field renormalization constant} introduced on (\ref{@55}). It is divergent for $\epsilon\rightarrow 0$ in $D=4-\epsilon$
dimensions.
In the critical limit $m^2\rightarrow 0$, the renormalization 
is power-like
$
\Phi_R (x)\rightarrow (\mu/\mu_0)^{-\eta/2} \Phi (x)$,
and  Eq.~(\ref{d12.130})
becomes
\begin{eqnarray} \label{d12.130P}
  \Gamma ^{(2)}_R (q) = - \left[
(q^2)^{2-\eta}  \mu^ \eta+ \left( \frac{m^2}{\mu ^2}\right)
         ^{\gamma} \mu ^{2}\right] .
\end{eqnarray}
The power $\eta$ is called the {\it anomalous dimension}
of the field $\Phi$. 

From (\ref{d12.130P}) we extract that the coherence length
of the system $\xi$ behaves like
\begin{eqnarray}
\xi
=\mu^{-1}(m^2)^{-\nu}.
\label{@Nuexpo}\end{eqnarray}
where
\begin{eqnarray}
\nu\equiv {\gamma/(2-\eta)}
\label{@Nuexpo}\end{eqnarray}
is the {\it critical exponent of the coherence length}.
\comment{see (\ref{rg.102})
\footnote{In Section
\ref{explsol}  we shall see
 that the coupling  strength governing the
behavior of all observable quantities in the limit $m^2\rightarrow 0$
is given by $\left( \hbar
g \sbar_D = \frac{2}{3 }\epsilon \right) $.
}
}

Another  power behavior is found for
$\Gamma ^{(4)}_R({\bf 0})$:
\comment{,
but here only in the limit $m^2 \rightarrow 0$
where $L^m \sim \log m^2/\mu ^2$, and thus
}%
\begin{eqnarray} \label{d12.132}\label{@93}
      \Gamma ^{(4)}_R({\bf 0}) &\dst \mathop{\rightarrow }_{m\to 0} &
             g \left(  1 + \frac{3}{4} \hbar g\mu^{-\epsilon} \sbar_{D}
              \log \frac{m^2}{\mu ^2} \right) + {\cal O} (m^2)\nonumber \\
         & = & g \left( \frac{m^2}{\mu ^2}\right) ^{\hbar \frac{3}{4}g\mu^{-\epsilon}
            \sbar_D}.
\end{eqnarray}
  Also this power behavior is measurable and defines the critical
  index $\beta $ via
the so-called scaling relation
\begin{eqnarray} \label{d12.133}
   \gamma -2\beta \equiv   \frac{3}{4} \hbar g\mu^{-\epsilon}\sbar_D,
\end{eqnarray}
so that 
\begin{eqnarray}
\beta=\frac12- \frac{1}{4} \hbar g\mu^{-\epsilon}\sbar_D.
\label{@betap}\label{@71}
\end{eqnarray}
The higher powers of $\Phi $
are accompanied by terms which are
 more and more
 singular in the limit $m\rightarrow 0$. 
We see from the
Feynman integrals in (\ref{d12.114}) that the diagrams
in (\ref{fluctint2})
behave  like $m^{4-\epsilon -n}$ for $m\rightarrow0$, and so do
the associated effective action terms $\Phi ^n$.

The coefficients of the dimensionless quantities
$(\Phi^{2}/\mu^{2-\epsilon})^\equiv \left(\tilde \Phi^{2}\right)^n$ or $(g\Phi^{2}/\mu^{2})^n\equiv \left(\hat \Phi^{2}\right)^n
$
have the general form
$(m^2/\mu^2)^{\gamma-2n\beta}$, so that the effective potential
can be written as 
\begin{eqnarray}
v(\Phi)=\mu^{4-\epsilon}\left(\frac{m^2}{\mu^2}\right)^\gamma 
\frac{\Phi^2}{\mu^{2-\epsilon}}f(x),
\label{@GenForM}\end{eqnarray}
with 
\begin{eqnarray}
x\equiv \left(\frac{m^2}{\mu^2}\right)^{-2\beta} 
\frac{\Phi^2}{\mu^{2-\epsilon}}
\equiv t^{-2\beta} \frac{\Phi^2}{\mu^{2-\epsilon}}=
 t^{-2\beta} 
{\tilde\Phi^2}
,
\label{@62B}\end{eqnarray}
\comment{or
\begin{eqnarray}
x&\equiv& 
\frac{\lambda}{ \sbar_D }\left(\frac{m^2}{\mu^2}\right)^{-2\beta}\tilde\Phi^2
,
\label{@}\end{eqnarray}
}
where we have abbreviated 
\begin{eqnarray}
t\equiv\frac{m^2}{\mu^2}.
\label{@}\end{eqnarray}
For small $\Phi$, the function $f(x)$ has a Taylor expansion in even powers 
of $\Phi$, corresponding to the diagrams in Eq.~(\ref{@38}):
\begin{eqnarray}
f(x)=
1+\sum_{n=1}^\infty f_n x^n=1+\sum_{n=1}^\infty f_n 
 \left(t^{-2n\beta}\tilde\Phi^2\right)^n.
\label{@60E}\label{@72}
\end{eqnarray}
If we start the sum at $n=-1$,
we also get the general form of the 
vacuum energy
\begin{eqnarray}
v(0)=\mu^{4-\epsilon} \left(\frac{m^2}{\mu^2}\right)^{\gamma+2\beta}.
\label{@}\end{eqnarray}
The exponent $\gamma+2\beta$ is equal to $D\nu$, where $\nu$ 
is the exponent defined in (\ref{@Nuexpo}).

By differentiating this energy twice with respect to $m^2$, we obtain 
the temperature behavior of the specific heat
\begin{eqnarray}
C\propto \left(\frac{m^2}{\mu^2}\right)^{D\nu-2}= \left(\frac{m^2}{\mu^2}\right)^{-\alpha}
\label{@}\end{eqnarray}
which yields the important 
critical exponent $\alpha=2-D\nu$ that governs the 
singularity of the famous $\lambda$-peak in superfluid helium at $T_c\approx 2.7$ Kelvin, the most accurately determined critical exponent in many-body systems
by a measurement in a satellite \cite{LIPA}.

{
We can also rewrite 
(\ref{@GenForM}) in the form
\begin{eqnarray}
v(\Phi)=\mu^{4-\epsilon}
\frac{ \sbar_D}{\lambda}
\left(\frac{m^2}{\mu^2}\right)^\gamma\frac{ g\Phi^2}{\mu^2} \bar f(y),
\label{@WIDOM10}\end{eqnarray}
where  
\begin{eqnarray}\!\!\!\!
\bar f(y)=
1+\sum_{n=1}^\infty\bar f_n y^n=1+\sum_{n=1}^\infty \bar f_n 
 \left(t^{-2\beta} \left(\frac{g\Phi^2}{\mu^{2}}\right)\right)^n\!\!.
\label{@60EE}\label{@74}
\end{eqnarray}
and
\begin{eqnarray}
y\equiv\frac{\lambda}{\sbar_D}x= g \mu^{-\epsilon}x=
\left(\frac{m^2}{\mu^2}\right)^{-2\beta} \left(\frac{g\Phi^2}{\mu^{2}}\right)
=t^{-2\beta}\hat \Phi^2
.
\label{@WIDOM2}\end{eqnarray}
}

In the limit $m^2\rightarrow 0$, the expansion 
(\ref{@60E}) and
(\ref{@60EE}) 
are divergent since the 
the coefficients grow like $n!$.
A sum can nevertheless be calculated with the technique
of Variational Perturbation Theory (VPT)
developed in the textbook \cite{KS22}.

\section{Massless Theory
and Widom Scaling}

Let us  evaluate the zero-mass limit
of $v(\Phi)$.
 Since $m^2$ always accompanies 
the coupling strength in the denominator, the limit $m^2\rightarrow 0$
is equivalent to the limit
$g\rightarrow \infty$, i.e., the strong-coupling limit.

The strong-coupling limit 
  deserves special attention. The theory in this limit is referred
  to the  {\it critical theory}.
\ins{critical,theory}%
\ins{theory,critical}%
 This name reflects the relevance  of this limit
  for the behavior of physical systems  at a critical
  temperature where fluctuations are of infinite range.

We shall see immediately that for large $y$,  $f(y)$ behaves like a 
pure power of $y$:
$f(y)\rightarrow y^{(\delta-1)/2}$, so that
\begin{eqnarray}
v(\Phi)\rightarrow \mu^{4-\epsilon}\frac1 {4!}\frac{ \sbar_D}{\lambda}\left(\frac{g\Phi^2}{\mu^2}\right)^{(\delta+1)/2}.
\label{@WEXPR}
\label{@82}
\end{eqnarray} 
Since without fluctuation corrections, $\delta=3$, this reduces 
properly to the mean-field potential
$g\Phi^4/4!$.

With this leading large $ \Phi$-behavior, we can rewrite the general  form 
(\ref{@WIDOM10}) of the potential
\comment{\begin{eqnarray}
v(\Phi)=\mu^2\left(\frac{m^2}{\mu^2}\right)f(x),~~~x\equiv g\Phi^2(\mu^2/m^2)^{2\beta}
\label{@}\end{eqnarray}
}
also as a so-called 
{\it Widom scaling expression\/}
depending on $y^{-1/2\beta}\propto  m^2/\Phi ^{1/\beta}$ \cite{WIDOM}:
\begin{eqnarray}
v(\Phi)\propto\Phi^{\delta+1}w(m^2/\Phi^{1/\beta}).
\label{@WIDOM}\end{eqnarray}
From this effective potential we may derive the general 
Widom form of the equation of state. After adding a source term $H\Phi$ 
and going to the extremum, we obtain 
 $H(\Phi)=\partial v(\Phi)/\partial \Phi$
with the 
general behavior
\begin{eqnarray}
H(\Phi)\propto\Phi^\delta h(m^2/\Phi^{1/\beta}).
\label{@}\end{eqnarray} 

Recalling (\ref{@WIDOM}), we 
expect the general form of the potential (\ref{@GenForM})
to be
\begin{eqnarray}
v(\Phi)\rightarrow \mu^{4-\epsilon}\frac1{4!}\frac{ \sbar_D}{\lambda}\left(\frac{g\Phi^2}{\mu^2}\right)^{(\delta+1)/2}
\hat w\left(\tau
\right)
,
\label{@WIDOM1N}\end{eqnarray}
where
 $\tau\equiv\lfrac {\left(\lfrac{m^2}{\mu^2}\right)}{(\Phi/\mu ^{1-\epsilon/2})^{1/\beta}}\equiv t/(\tilde \Phi^2{})^{1/\beta}$.
\comment{
the proper dimensions, it is equal to
\begin{eqnarray}
v(\Phi)=\mu^{4-\epsilon}(\Phi/\mu^{1-\epsilon/2})^{\delta+1}\hat w\left(\frac{m^2/\mu^2
}{(\Phi/\mu ^{1-\epsilon/2})^{1/\beta}}\right).
\label{@WIDOM1N}\end{eqnarray}
}

For small $m^2$, $\hat w(\tau)$ has a series expansion
in powers of $\tau^{\omega\nu}$:
\begin{equation}
\hat w(\tau)\!=\!1\!+\!c_1\tau^{\omega\nu}
+\!c_2\tau^{2\omega\nu}+ \dots),
\label{@WE}\end{equation}
or since $t=\mu\xi^{-1/\nu}$:
\begin{equation}
\hat w(\tau)\!=\!1\!+\!\bar c_1\xi^{-\omega}\Phi^{-\omega\nu/\beta}\!+\!\bar c_2\xi^{-2\omega}
\Phi^{-2\omega\nu/\beta}+ \dots).
\label{@WE}\end{equation}
Here $\omega$ is the Wegner exponent \cite{WEGNER}
that governs the approach to scaling.
Its numerical value is close to $0.8$ \cite{REMKIN}.

Differentiating (\ref{@WIDOM1N}) with respect to $\Phi$ yields
the following leading contribution to $H$:
\begin{eqnarray}
H&=&\partial_\Phi v(\Phi)=\frac{\delta+1}{4!}\mu^2\left(\frac{g\Phi^ 2}{\mu^2}\right)^{(\delta-1)/2}\nonumber \\&=&
\frac{\delta+1}{4!}\mu^2\left(\frac{\lambda\tilde\Phi^ 2}
{\sbar_D}\right)^{(\delta-1)/2}
,
\label{@HEQA}\end{eqnarray}
where $\delta-1=
\gamma/\beta$.

Note that the effective potential remains
finite for $m=0$. Then, $v(\Phi)$ becomes
\def\ns{\hspace{-1pt}}
\begin{eqnarray} \label{d12.136}\hspace{-1pt}\!\!\!\!\!\!\!\!\!\!\ns
 \!\! v(\hspace{-1pt}\Phi\ns)\!\!\!\! & = &\!\!\! \frac{g}{4!} \Phi ^4\hspace{-1pt}\!+\! \frac{
\hbar}{2}\!\ns \int
\!\!         \frac{d^Dq}{(2\pi )^D}\! \log \left(\ns \! 1\!\hspace{-1pt} +\! \frac{g}{2}\hspace{-1pt}
          \frac{\Phi ^2}{q^2}\ns\!\right)\!\! +\! \frac{\hbar g^2\mu ^{-\epsilon }}{16}
          \frac{  \sbar_D}{\epsilon }\hspace{-1pt} \Phi ^4\!\!\hspace{1pt}.
\end{eqnarray}
It 
displays an important feature: When
expanding
the logarithm
in powers of $\Phi $, the
 expansion terms
 correspond to  increasingly  divergent
Feynman integrals $$\int \frac{d^Dq}{(2\pi )^D} \frac{1}{(q^2)^n}.$$
Contrary
 to the previously regularized divergencies  coming from the  large-$q^2$ regime,
 these divergencies are due to the $q=0$ -singularity
of the massless propagators $G_0(q)=i/q^2$. This means that they are 
IR-singularities.
Let
us verify that the effective potential remains
indeed finite for $m=0$.%
\comment{ Then, $v(\Phi)$ becomes
\begin{eqnarray} \label{d12.136}\!\!\!\!\!\!\!\!\ns
 \!\! v(\hspace{-1pt}\Phi\ns)\!\!\!\! & = &\!\!\! \frac{g}{4!} \Phi ^4\!+\! \frac{
\hbar}{2}\!\ns \int
\!\!         \frac{d^Dq}{(2\pi )^D}\! \log \left(\ns \! 1\! +\! \frac{g}{2}
          \frac{\Phi ^2}{q^2}\ns\!\right)\!\! +\! \frac{\hbar g^2\mu ^{-\epsilon }}{16}
          \frac{  \sbar_D}{\epsilon } \Phi ^4\!\!.
\end{eqnarray}
It 
displays an important feature: When
expanding
the logarithm
in powers of $\Phi $, the
 expansion terms
 correspond to  increasingly  divergent
Feynman integrals $$\int \frac{d^Dq}{(2\pi )^D} \frac{1}{(q^2)^n}.$$
Contrary
 to the previously regularized divergencies  coming from the  large-$q^2$ regime,
 these divergencies are due to the $q=0$ -singularity
of the massless propagators $G_0(q)=i/q^2$. This means that they are 
IR-singularities.%
\comment{
Such
infrared (IR)
divergences were encountered before in quantum electrodynamics in
Chapter~\ref{QED}.\ins{infrared problems}\ins{infrared singularities}
}%
It is a pleasant feature of the sum over all these
terms
that the final result has lost the IR singularities
 in favor of a weak logarithmic singularity,
thus removing the IR problem.
 The only singularities which
  remain lie in the ultraviolet, and these are taken care of
 by the  counter terms.
}%
\comment{
  In $4-\epsilon $ dimensions, this behaves like
\begin{eqnarray} \label{d12.139}
  \sbar_D \left[ -1/\epsilon + {\cal O} (\epsilon )
            \right] \frac{1}{2} \left( 1 +\frac{ \epsilon}{4} \right)
        \!    \left( m^2\right) ^{\frac{2-\epsilon }{2}}\!\!\!,
\end{eqnarray}
}
Performing the momentum integral in (\ref{d12.136})
the potential becomes
\begin{eqnarray} \label{d12.140}
\!\!\!\!\!  v(\Phi )\!\!\! & = & \!\!\!\frac{g}{4!} \Phi ^4\! +\! \frac{
\hbar}{4}
                 \sbar_D \left( \!-\frac{1}{\epsilon }
\!
 \right) \left( \frac{g}{2}\Phi ^2
                  \right) ^{2-\frac{\epsilon }{2}}\!
            \!  \!  +\! \frac{
\hbar g^2}{16} \sbar_D \mu ^{-\epsilon }
                  \frac{1}{\epsilon } \Phi ^4 \!.\!\!\!\!\nonumber \\
\end{eqnarray}
With the goal of expanding this for small 
$\epsilon $-expansion, we divide the coupling
   constant and field by a scale parameter involving
   $\mu $.
   Then $g\mu ^{-\epsilon }$ and $\Phi /\mu ^{1-\epsilon /2}$
   are dimensionless quantities, in terms of which
\begin{eqnarray} \label{d12.141}
  &&\!\!\!\!\!\!\!\!\!\!\!\!\!\!\!\!\!\!\!\!\! v(\Phi ) \! =\mu ^{4-\epsilon } \left[ \frac{g\mu ^{-\epsilon }}
                  {4!} \left( \frac{\Phi }{\mu ^{1-\epsilon /2}}\right) ^4
                  \!+\! \frac{\hbar}{4} \sbar_D \left(\! - \frac{1}{\epsilon }\!
\right)
\right.
\nonumber \\&&\!\! \left.\!\!\!\!\!\!\!\!\!~\hspace{0pt}\times\!
                   \left( \frac{g\mu ^{-\epsilon }}{2}
                 \frac{\Phi ^2}{\mu ^{2-\epsilon }}\right) ^{2-\frac{\epsilon }
                 {2}}\right]
\!\!+\! \frac{\hbar g^2}{16} \frac{\mu ^{-2\epsilon }}
                       {\epsilon } \sbar_D \!\left( \frac{\Phi }
                       {\mu ^{1-\epsilon /2}}\right) ^4\!\!.
\end{eqnarray}
If we use the dimensionless coupling constant
\begin{eqnarray}
 \lambda \equiv  \sbar_D \hbar g \mu ^{-\epsilon}
\label{@RCC}\end{eqnarray}
and the reduced field $\tilde{\Phi }\equiv\Phi /\mu ^{1-\epsilon /2}$
as new variables, then
\begin{eqnarray} \label{d12.142}\!\!\!\!\!\!\!\!\!\!\!\!\!
&&\!\!\!\! \hspace{-4em}  v(\Phi ) \! =\!  \frac{\mu ^{4-\epsilon }}{\hbar\sbar_D}
                 \left[ \frac{\lambda }{4!} \tilde{\Phi }^4
                 + \frac{1}{4} \left( \!-\frac{1}{\epsilon }
 \!              
\right)\! \left( \frac{\lambda\tilde{\Phi }^2 }{2}
                 \right) ^2 \nonumber \right.\\&&\left.\times \left( 1 - \frac{\epsilon }{2} \log
                   \frac{\lambda \tilde{\Phi }^2}{2} \right)
                + \frac{\lambda ^2}{16\epsilon }
                      \tilde{\Phi }^4
                   \right\}. 
\label{@89}
\end{eqnarray}
To zeroth order in $\epsilon $,
the prefactor is  equal to $\mu ^4\, 8\pi ^2$.
Thus the massless limit of the effective potential is well
defined in $D= 4$ dimensions. There is, however, a special feature:
 The finiteness  is achieved at the expense of
 introducing
 the extra parameter $\mu $.

The most important property of the critical potential
is that it cannot be expanded in  integer powers of $\Phi^2 $.
Instead, the expression (\ref{d12.142}) can be rewritten, 
correctly up to order $\lambda ^2$,  as
\begin{eqnarray} \label{d12.143}
 v(\Phi )\!\!\!\! & = &\!\!\!\! \frac{\mu ^{4-\epsilon }}{\sbar_D} \frac{1}{6}
                \left\{ \left( \frac{\lambda }{2} 
 \tilde{\Phi }^2\right) ^2\!\!
                + \frac{3}{8}\! \left(\! \frac{\lambda }{2}\tilde{\Phi }^2\!
                \right) ^2 \log \frac{\lambda }{2} \tilde{\Phi }^2\!
                \right\} \nonumber \\
            & = & \frac{\mu ^{4-\epsilon }}{\sbar_D} \frac{1}{6\lambda }
                  \left( \frac{\lambda }{2} 
\tilde{\Phi }^2\right) ^{2+
                  \frac{3}{4}\lambda } + {\cal O}(\lambda ^3).
\label{@91}
\end{eqnarray}
This is the typical 
power behavior of the critical interaction.
The power of $\tilde \Phi $
defines the {\it critical exponent of the interaction} $1 + \delta $,
so that up to the first order in the 
coupling strength:
\begin{eqnarray}
\delta  = 3 + \frac{3}{2}\lambda .
\label{@108AB}
\end{eqnarray}
At the mean-field level,
$\delta$ is equal to $3$

\comment{
The expansions in powers of $\epsilon $
has an important property which
has the direct applications  in the
description of the strong-coupling limit, i.e., 
in critical phenomena. For small
$g $, Eq.~(\ref{d12.123})  can be rewritten as
\begin{eqnarray} \label{d12.129}
  \Gamma _R^{(2)} =  \left( q^2 + m^2 +
           \frac{ \hbar}{4} g\mu^{-\epsilon}\sbar_D \log \frac{m^2}{\mu ^2}\right)
\end{eqnarray}
which is,
to the same order in $g$,  equal to
\begin{eqnarray} \label{d12.130}
  \Gamma ^{(2)}_R (q) =  \left[ q^2 + \left( \frac{m^2}{\mu ^2}\right)
         ^{1+ \frac{\hbar}{4}g\mu^{-\epsilon} \sbar_D} \mu ^2\right] .
\end{eqnarray}
This means that the vertex function at $q=0$
has a mass term 
that depends on the mass $m$ of the $\phi$-field
via a power law:
\begin{eqnarray} \label{d12.131}
 \label{@Form52}
  \Gamma _R^{(2)} (0) =  \left( \frac{m ^2}{\mu ^2}\right) ^\gamma
           \mu ^2.
\end{eqnarray}
The power 
$\gamma$
depends in 
the coupling strength $g$ like
\begin{eqnarray} \label{d12.132}
  \gamma  = 1 + \hbar \frac{g\mu^{-\epsilon}}{4} \sbar_D.
\end{eqnarray}
The important point is that this power $ \gamma$ is measurable
as an experimental
quantity called the \iem{susceptibility}.
It is the
so-called {\it critical exponent}
\ins{critical,exponent}%
\ins{exponent;critical}%
 of the
susceptibility.
}
\comment
{
\section{Widom Scaling}
%
Let us first  evaluate the zero-mass limit
of $v(\Phi)$.
 Since $m^2$ always accompanies 
the coupling strength in the denominator, the limit $m^2\rightarrow 0$
is equivalent to the limit
$g\rightarrow \infty$, i.e., the strong-coupling limit.
The strong-coupling limit 
  deserves special attention. The theory in this limit is referred
  to the  {\it critical theory}.
\ins{critical,theory}%
\ins{theory,critical}%
 This name reflects the relevance  of this limit
  for the behavior of physical systems  at a critical
  temperature where fluctuations are of infinite range.
We shall see immediately that for large $y$,  $f(y)$ behaves like a 
pure power of $y$:
$f(y)\rightarrow y^{(\delta-1)/2}$, so that
\begin{eqnarray}
v(\Phi)\rightarrow \mu^{4-\epsilon}\frac1 {4!}\frac{ \sbar_D}{\lambda}\left(\frac{g\Phi^2}{\mu^2}\right)^{(\delta+1)/2}.
\label{@WEXPR}\end{eqnarray} 
Since without fluctuation corrections, $\delta=3$, this reduces 
properly to the mean-field potential
$g\Phi^4/4!$.
With this leading large $ y$-behavior, we can rewrite the general  form 
(\ref{@WIDOM10}) of the potential
\comment{\begin{eqnarray}
v(\Phi)=\mu^2\left(\frac{m^2}{\mu^2}\right)f(x),~~~x\equiv g\Phi^2(\mu^2/m^2)^{2\beta}
\label{@}\end{eqnarray}
}
can also be rewritten as a so-called 
{\it Widom scaling expression\/}
depending on $x^{-1/2\beta}\propto  m^2/\Phi ^{1/\beta}$:
\begin{eqnarray}
v(\Phi)\propto\Phi^{\delta+1}w(m^2/\Phi^{1/\beta}).
\label{@WIDOM}\end{eqnarray}
From this effective potential we may derive the general 
Widom form of the equation of state. After adding a source term $H\Phi$ 
and going to the extremum, we obtain 
 $H(\Phi)=\partial v(\Phi)/\partial \Phi$
with the 
general behavior
\begin{eqnarray}
H(\Phi)\propto\Phi^\delta h(m^2/\Phi^{1/\beta}).
\label{@}\end{eqnarray} 
%
%
Recalling (\ref{@WIDOM}), we 
expect the general form of the potential (\ref{@GenForM})
to be
\begin{eqnarray}
v(\Phi)\rightarrow \mu^{4-\epsilon}\frac1{4!}\frac{ \sbar_D}{\lambda}\left(\frac{g\Phi^2}{\mu^2}\right)^{(\delta+1)/2}
\hat w\left(\tau
\right)
,
\label{@WIDOM1}\end{eqnarray}
where
 $\tau\equiv\lfrac {\left(\lfrac{m^2}{\mu^2}\right)}{(\Phi/\mu ^{1-\epsilon/2})^{1/\beta}}\equiv \hat m^2/(\tilde \Phi^2{})^{1/\beta}$.
\comment{
the proper dimensions, it is equal to
\begin{eqnarray}
v(\Phi)=\mu^{4-\epsilon}(\Phi/\mu^{1-\epsilon/2})^{\delta+1}\hat w\left(\frac{m^2/\mu^2
}{(\Phi/\mu ^{1-\epsilon/2})^{1/\beta}}\right).
\label{@WIDOM1}\end{eqnarray}
}
For small $m^2$, $\hat w(\tau)$ has a power series expansion
\begin{equation}
\hat w(\tau)\!=\!1\!+\!c_1\xi^{-\omega}\Phi^{-\omega\nu/\beta}\!+\!c_2\xi^{-2\omega}
\Phi^{-2\omega\nu/\beta}+ \dots).
\label{@WE}\end{equation}
Here $\omega$ is the Wegner exponent 
that governs the approach to scaling   \cite{REMKIN}.
Its numerical value is close to $0.8$.
Differentiating (\ref{@WIDOM1}) with respect to $\Phi$ yields
the following leading contribution to $H$:
\begin{eqnarray}
H&=&\partial_\Phi v(\Phi)=\frac{\delta+1}{4!}\mu^2\left(\frac{g\Phi^ 2}{\mu^2}\right)^{(\delta-1)/2}\nonumber \\&=&
\frac{\delta+1}{4!}\mu^2\left(\frac{\lambda\tilde\Phi^ 2}
{\sbar_D}\right)^{(\delta-1)/2}
,
\label{@HEQA}\end{eqnarray}
where 
\begin{eqnarray}
\delta=1+
\gamma/\beta.
\label{@108A}\end{eqnarray}
}

\section{Critical Coupling Strength}

What is 
the coupling strength $\lambda$ in the critical regime?
The counter term proportional to $\Phi^4$ in
Eq.~(\ref{d12.136}) implies that the use of the 
renormalized 
coupling constant in the subsequent subtracted expressions.
This corresponds to the use of a
bare coupling constant $g_B$ 
instead of $g$ in the original action
(\ref{d12.102B}). The relation between the two 
to this one-loop order is
\begin{eqnarray}
g_B=g+\frac 3 2\hbar g^2 \mu^{-\epsilon}
          \frac{  \sbar_D}{\epsilon }+\dots~.
\label{@BARE}\end{eqnarray} 
For a given bare interaction strength $g_B$, the renormalized coupling
depends on the parameter $\mu$ chosen for the renormalization
procedure.
Equivalently we may imagine having defined the field theory 
on a fine spatial lattice with a specific
small lattice spacing $a\propto 1/\mu$,
and the renormalized coupling constant 
will
depend on the choice of $a$.

If we sum an infinite chain of such corrections,
we obtain a geometric sum 
that is an expansion of the equation
\begin{eqnarray}
g_B=\frac{g}{1-\dst\frac 3 2 \hbar g \mu^{-\epsilon}
          \frac{  \sbar_D}{\epsilon }},
\label{@107}\end{eqnarray}
which has (\ref{@BARE}) as its first expansion term.
Equivalently we may write
\begin{eqnarray}
\frac{1}{ 
\hbar g_B\mu^{-\epsilon}}=\frac{1}{
\hbar
g\mu^{-\epsilon}}-\frac 3 2 
          \frac{  \sbar_D}{\epsilon }+\dots~.
\label{@}\end{eqnarray}
In this equation, we can go to the
strong-coupling limit $g_B\rightarrow \infty$
by taking $\mu $ to the
 critical limit
 $ \mu\rightarrow 0$, where we
 find that the
renormalized coupling has a finite value
\begin{eqnarray}
\frac{1}{\hbar g\mu^{-\epsilon}}
=
\frac 3 2 
          \frac{  \sbar_D}{\epsilon }
  +\dots~.
\label{@110}\end{eqnarray} 
For the
dimensionless coupling constant
(\ref{@RCC}), this amounts
to the
strong-coupling limit
\begin{eqnarray}
\lambda\equiv
{\hbar g\mu^{-\epsilon}} \sbar_D\rightarrow\frac 2 3 \epsilon
+\dots~,
\label{@GSTAR}\end{eqnarray} 
with the omitted terms being of higher order in $\epsilon$.
The approach to this limit starting from small coupling 
is obtained 
from (\ref{@107})
to be
\begin{eqnarray}
\hbar g \mu^ {-\epsilon}=\frac1{\dst\frac1{\hbar g_B \mu ^{-\epsilon}}+
\dst\frac 3 2
          \frac{  \sbar_D}{\epsilon }
}.
\label{@107b}\end{eqnarray}

At small bare coupling constant,
this starts out with 
the renormalized expression  that is determined by the 
$s$-wave scattering length
$
\hbar g \mu^ {-\epsilon}=6 \times 4\pi \hbar ^2 a_s/M$.
In the strong-coupling limit of $g_B$ 
or the critical  limit 
$\mu\rightarrow 0$, the
value
(\ref{@110}) is reached.

For the critical exponents 
 $\gamma,\beta,\delta$ in 
(\ref{@65}),
(\ref{@71}),
(\ref{@108AB}),
the strong-coupling limits are
\begin{eqnarray}
\gamma=1+\frac16 \epsilon,~~
\beta=\frac12-\frac16 \epsilon,~~
\delta=3+\epsilon.
\label{@CREXP}\end{eqnarray}
They are  approached for finite $g_B$
like
\begin{eqnarray}
\gamma=1+\frac1 4
{\hbar g \mu ^{-\epsilon}}
,~~
\beta=\frac12-\frac1 4
{\hbar g \mu ^{-\epsilon}}
,
\label{@CREXP3}\end{eqnarray}
with the just discussed $g_B$-behavior of $g$.

If the
omitted terms
in (\ref{@GSTAR}) are calculated 
for all $N$ and to higher loop orders 
one finds $\lambda=2 g^*$, 
where $g^*$
is the strong-coupling limit $g_B\rightarrow \infty$
of the series
in Eq. (15.18) in the textbook \cite{KS22}].
The other critical
exponents may be obtained from the  
 $\bar g_B 
\rightarrow \infty$ -limit of similar expansions
for $\nu, \eta$ into
$\alpha=2-\nu D$, $\beta=\nu(D-2+\eta)$,
$\gamma=\nu(2-\eta)$,
$\delta=(D+2-\eta)/(D-2+\eta)$.
These can be extracted from Chapter 15 in Ref.  \cite{KS22}.

All these series are divergent, but can be resummed for $\epsilon=1$ and $\bar g_B 
\rightarrow \infty$.
The series for $\bar g$ 
yields for $N=2$ the value
$\bar g\rightarrow
g^*\approx 0.503$ (see Fig. 17.1 in \cite{KS22}).
The typical dependence of $\bar g$ on $\bar g_B$ is plotted in Fig. 20.8 of Ref. \cite{KS22}.
For $\nu$ and $\eta$ the plots are shown in Figs. 20.9 and 20.10, and we leave it to the reader to compose from these the 
dependence 
of $\alpha,\beta,\delta$ as functions of 
$\bar g_B$.

It should be pointed out that the
potential 
in the first line of (\ref{d12.143}) 
has another minimum away
from the origin at
\begin{eqnarray} \label{d12.144}
  \frac{\lambda }{3!}\tilde  \Phi ^3 + \frac{\lambda ^2}{8}
           \tilde{\Phi }^3 \left( \log \frac{\lambda }{2}
           \tilde{\Phi }^2 - \frac{1}{2}\right) +
           \frac{\lambda ^2}{16} \tilde{\Phi }^3 =0.
\end{eqnarray}
This is solved by
\begin{eqnarray} \label{d12.145}
  \lambda \,  \log \frac{\lambda }{2} \tilde{\Phi }^2
   = - \frac{4}{3}
\end{eqnarray}
or
\begin{eqnarray} \label{d12.146}
        \tilde{\Phi }^2 = \frac{2}{\lambda } e^{-4}{3\lambda }.
\end{eqnarray}
However such a solution
found for small $ \lambda $
cannot be trusted.
The higher loops  to be discussed later and
neglected up to this point will produce more powers
of $\lambda \log (\lfrac{\lambda \tilde{\Phi }^2}{2}) $, and
the series cannot be expected to converge at such a large $ \lambda $.
 As a matter of fact, the approximate exponentiation
performed in 
(\ref{@91}) 
does not show this minimum and will be seen,
via the methods to be described later,
to be the correct analytic form of the potential to all orders
in $\lambda $ for small enough $\epsilon $ and $ \lambda $.

If we want to apply the formalism to a Bose-Einstein-condensate
we must discuss the case of 
a general
O($N$)-symmetric
version of the effective potential based on the action
(\ref{d12.109}), and insert into it the number $N=2$. 
The equation for the bare coupling constant
is then 
\begin{eqnarray}
g_B=g+\frac {N+8}6\hbar g^2 \mu^{-\epsilon}
          \frac{  \sbar_D}{\epsilon }+\dots~,
\label{@Form89}\end{eqnarray} 
rather than (\ref{@BARE}), so that the strong-coupling limit
(\ref{@GSTAR})
becomes
\begin{eqnarray}
\lambda\equiv
{ \sbar_D \hbar g\mu^{-\epsilon}}\rightarrow\frac 6 {N+8} \epsilon
+\dots~,
\label{@GSTAR1}\end{eqnarray} 
with the other critical exponents
(\ref{@CREXP}):
\begin{eqnarray}\!\!\!\!\!\!
\gamma\!=\!1\!+\!\frac{N+2}{2(N\!
+\!8)} \epsilon,~
\beta\!=\!\frac12\!-\!\frac{3}{2(N\!+\!8)} \epsilon,~
\delta\!=\!3\!+\!\frac9{N\!+\!8}\epsilon.\label{@Form91}
\label{@CREXPP}\end{eqnarray}

For the coupling constant $g_S$ defined in (\ref{@gSg})
the strong-coupling limit
(\ref{@GSTAR1})
reads
\begin{eqnarray}
{g_S}\rightarrow 2\, \frac1{4!}\frac{\lambda}{\sbar_D}\mu^\epsilon=\frac{(4\pi)^2}{4!}\lambda\,
\mu^\epsilon=\frac{(4\pi)^2}{12}g^*\,
\mu^\epsilon,
\label{@}\end{eqnarray}
with $g^*\approx 0.503 $ for $N=2$.

If we carry the loop expansion to higher order in $\hbar$, we find for 
the renormalized $\bar g =\hbar g_B\mu^{-\epsilon}/2\sbar_D$
 as function of the bare coupling $\bar g_B =\hbar g_B\mu^{-\epsilon}/2\sbar_D$ the
perturbation
expansion [see Eq.~(15.18) in Ref.~\cite{KS22}. See there also
the  higher expansion terms]
\newcommand{\hmbx}[1]{\hspace*{-3.5mm}\mbox{\small$#1$}}
\newcommand{\mbx}[1]{\mbox{\small$#1$}}
\def\ep{\epsilon}
\comment{
\begin{eqnarray}
\lefteqn{\hspace*{0.cm}\mbx{
g^*(\ep) =  ~~ 
{{3\,  \ep}\over {8 + N}} +
   {{9\,{{\ep}^2} }\over {{{\left( 8 + N \right) }^3}}}
                             \left( 14 + 3\,N \right) } }
\nonumber\\
&\mbx{+{{\ep^3}\over {\left( 8 + N \right)^5}} }
&\hmbx{\,\bigl[
       \frac{3}{8}\,\left( 4544 + 1760\,N + 110\,{N^2} - 33\,{N^3} \right) 
 }
\nonumber\\
&\mbx{\phantom{+{{\ep^3}\over {\left( 8 + N \right)^5}} }}
&\hmbx{\,~
-
        \zeta (3)\cdot 36\,\left( 8 + N \right)\left( 22 + 5\,N \right)
       \bigr] }
\nonumber\\
&\mbx{+{{\ep^4}\over {{\left( 8 + N \right) }^7}} }
&\hmbx{\,\bigl[
     \frac{1}{16} (529792+309312\,N + 52784\,{N^2}- 5584\,{N^3}
 }
\nonumber\\
&&\hmbx{~~
    - 2670\,{N^4}-5\,{N^5}) }
\nonumber\\
&&\hmbx{~~\hspace{-3em}
    + \zeta (3)(8+N)\cdot 6
        \left( -9064 - 3796\,N - 82\,{N^2} + 63\,{N^3} \right) }
\nonumber\\
&&\hmbx{~~\hspace{-3em}
    - \zeta (4)(8+N)^3 \cdot 18\left( 22 + 5\,N \right) }
\nonumber\\
&&\hmbx{~~\hspace{-3em}
    + \zeta (5)(8+N)^2\cdot 120 \left( 186 + 55\,N + 2\,{N^2} \right)
       \bigr] }\!+\dots~\!.
\comment{\nonumber\\
&\mbx{+{{{\ep}^5}\over {\left( 8 + N \right)^9}} }
&\hmbx{\bigl[\frac{3}{256}(\hspace*{-2mm}
         \begin{array}[t]{l}
          \mbx{-21159936-8425472\,N+3595520\,{N^2}+758144\,{N^3} }
           \\[-1.0mm]
          \mbx{-625104\,{N^4}-179408\,{N^5}-1262\,{N^6}-13\,{N^7})}
         \end{array} }
\nonumber\\
&&\hmbx{~~
    + \zeta (3) (8+N)\cdot\frac{3}{16}(\hspace*{-2mm}
         \begin{array}[t]{l}
           \mbx{-15131136-8873728\,N-890208\,{N^2}}
           \\[-1.0mm]
           \mbx{+310248\,{N^3}+45592\,{N^4}-1104\,{N^5}+9\,{N^6})}
         \end{array} }
\nonumber\\
&&\hmbx{~~
    +{\zeta (3)}^2 (8+N)^2\cdot 9
        \left( 43584 + 24848\,N + 3626\,{N^2}+107\,{N^3}+6\,{N^4}\right) }
\nonumber\\
&&\hmbx{~~
    + \zeta (4)(8+N)^3\cdot\frac{27}{8}
        \left( -8448 - 3524\,N - 52\,{N^2} + 63\,{N^3} \right)  }
\nonumber\\
&&\hmbx{~~
    + \zeta (5)(8+N)^2\cdot 3
        \left( 554208+255188\,N+12246\,{N^2}-5586\,{N^3}-305\,{N^4} \right)}
\nonumber\\
&&\hmbx{~~
    + \zeta (6)(8+N)^4\cdot\frac{225}{2}
        \left( 186 + 55\,N + 2\,{N^2} \right) }
\nonumber\\
&&\hmbx{~~
    - \zeta (7)(8+N)^3\cdot 1323
        \left( 526 + 189\,N + 14\,{N^2} \right)
       \bigr] }.
}
\label{@epsgsta}
\end{eqnarray} 
}
\begin{eqnarray} 
\frac{\bar g}{\bar g_B}\!\!\!&=&\!\!\!\textstyle  1
 -  \bar{g}_B\,{{8 + N }\over {3}}\,\ep^{-1}
 +\, {\bar{g}_B^2}\,\left\{ {\left( 8 + N \right)^2 \over {9 }}{1\over\ep^{2}} +
                        { 14 + 3\,N\over {6}} {{1}\over \ep^{}}
                  \right\}
\nonumber\\&&\hspace{.5mm}\textstyle
 + ~{\bar{g}_B^3}\left\{
            -{\left( 8 + N \right)^3 \over 27}{{1}\over{\ep^{3}}} -
          4\,{\left( 8 + N \right) \left( 14 + 3N \right)\over{27}}
{{1}\over{\ep^{2}}}
  \right.\label{@grenofgbar1}
\\&&\textstyle\hspace*{0.5cm} \left.
           -  {\left[ 2960 + 922\,N + 33\,{N^2} +
                 \left( 2112 + 480\,N\right) \zeta (3)
              \right]\over{648}}\,{{1}\over{\ep^{}}}
                  \right\}+\dots~.\nonumber 
\end{eqnarray} 
\comment{
Resumming this via VPT one obtains the 
curve shown in Fig. \ref{plgrfl},  
together  with the divergent weak-coupling
partial sums and the convergent strong-coupling partial sums.
\begin{figure}[htbp]
~\\[-.5cm]
\begin{picture}(94.49,157.135)
\put(-60,-40){\includegraphics[width=3.65cm]{plgrfl.eps}}
\put(55,55){ $\bar g$}
\put(125,10){ $\log_{10}\bar g_B$}
\end{picture}
\caption[
Logarithmic plot
of variational perturbation result
for
 expansion
(\protect\ref{gfg-0})  of renormalized coupling constant
$\bar g(\bar g_B)$
with $N=1$
]{
Logarithmic plot
of variational perturbation result
for expansion
(\ref{@grenofgbar1})
of renormalized coupling constant
$\bar g(\bar g_B)$
 with $N=1$
as a function of the bare coupling constant
$ \bar g_B$.
The small-$\bar g_B$ regime shows truncated
divergent perturbation expansions; the
large-$\bar g_B$ regime shows the
truncated
strong-coupling expansions.
Dash lengths increase with
 orders.
}
\label{plgrfl}\end{figure}
}

Given this dependence of $\bar g$
on the bare coupling constant  $\bar g_B$,
we find for $\gamma$ the expansion
\begin{eqnarray}
\lefteqn{\hspace*{1pt}\!\!~ ~~~\hspace*{-2em}\mbx{\gamma(\bar g_B)
\!=\! \frac{\bar g}{6}(N\!+\!2)
-5\frac{\bar g^2}{36}(N\!+\!2)
+\frac{\bar g^3}{72}(N\!+\!2)\bigl(5N+37\bigr)} }
\nonumber \\
&\mbx{\!\!~~~~-\frac{\bar g^4}{15552}(N\!+\!2) ~}
 &\hmbx{\bigl[-N^2+7578N+31060 }
\label{analgamm2} \\
&
 &\hspace*{-6em}\hmbx{~~~\!\!\!\hspace{.1pt}+48\zeta(3)(3N^2\!+\!10N\!+\!68) }
+288\zeta(4)(5N\!+\!22)\bigr]\!+\!\dots\! \hspace{-1pt}
\nonumber 
\comment{
\nonumber \\
&~~~~~~~~\mbx{+\frac{\bar g^5}{373248}(N\!+\!2) }
 &\hmbx{\bigl[21N^3+45254N^2+1077120N+3166528 }
\nonumber \\
&
 &\hmbx{+\,\zeta(3)\cdot48(17N^3+940N^2+8208N+31848) }
\nonumber \\
&
 &\hmbx{-\,\zeta^2(3)\cdot768(2N^2+145N+582) }
\nonumber \\
&
 &\hmbx{+\,\zeta(4)\cdot288(-3N^3+29N^2+816N\!+\!2668) }
\nonumber \\
&
 &\hmbx{+\,\zeta(5)\cdot768(-5N^2+14N+72) }
\nonumber \\
&
 &\hmbx{+\,\zeta(6)\cdot9600(2N^2+55N+186)\bigr] }
}~,\!\!\!\!\!\!
\label{@}\end{eqnarray}
where $\bar g$ is replaced by
(\ref{@grenofgbar1}). 

The expression 
$(m^2/\mu^2)^\gamma$ in the two-point function
can be replaced 
by an expansion of $m^2/m_B^2$ 
in powers of $\bar g_B$ that can be taken from Eq.~(15.15) in 
\cite{KS22},
again
with $\bar g$  replaced by
(\ref{@grenofgbar1}). 
This expansion can be resummed by Variational Perturbation Theory 
to obtain a curve
of the type in Fig.~20.8--20.10.
This permits us to relate $t=m^2/\mu^2$ directly to $\mu^{-\epsilon}g_B$.

{
\section{Resumming the Effective Potential}
According to Eq.~(\ref{@WIDOM10}),
the effective action 
below in the condensed phase with negative $m^2$
has the general form
\begin{eqnarray}
\frac{v(\Phi)}{\mu^{4-\epsilon}}=
t^{\gamma}\frac{\Phi^2}{\mu^{2-\epsilon}}
(\bar f_0+\bar f_1 y+\dots ~),
\label{@104}\end{eqnarray}
where 
\begin{eqnarray}
y=t^{-2\beta}g\Phi^2/\mu^{2}=t^{-2\beta}\hat\Phi^2.
\label{@123a}\end{eqnarray}
From Eq.~%
(\ref{@74} 
 we determine $\bar f_0=\bar f_1=1$.
For small $t$, $y$ becomes large, and we must convert the small-$y$ expansion
into a large-$y$ expansion.

Near the strong-coupling limit,
the Widom function (\ref{@WIDOM}) has an expansion
in powers of $( m^2)^{\omega/\nu}\propto \xi^{-\omega}$ which  
contains 
the Wegner
critical exponent
$\omega\approx 0.8$ governing the
{\it approach to scaling\/}  \cite{REMKIN}.
The general expansion for strong couplings is
\begin{eqnarray}
\frac{v(\Phi)}{\mu^{4-\epsilon}}\!\!\!\!&=&\!\!\!t^{\gamma}\!\frac{\Phi^2}{\mu^{2-\epsilon}}(t^{-2\beta}\hat\Phi^2)^{(\delta-1)/2}\!\ns
\left(\!b_0\!\hspace{1pt}+\!\!\sum_{m=1}^\infty\!\frac{b_m}{(t^{-2\beta}\hat\Phi^2)^{m\omega\nu/2\beta}}
\!\!\right)\!\ns
,\nonumber \\
\label{@105}\label{@116}\end{eqnarray}
We may derive this 
from the rules of VPT in 
 \cite{KS22}. First we rewrite a variational ansatz for
the right-hand side of 
Eq.~(\ref{@104}) with the help of a dummy parameter $\kappa=1$ as
\begin{eqnarray}
w_N
=\mu^ 2 t^\gamma\Phi^2 \kappa^p\left(\bar f_0 +\bar f_1 \frac{y}{\kappa^q}+\dots~\right)
.
\label{@106}\end{eqnarray}
Next we exchange $\kappa^p$ by the identical expression $ \sqrt{K^2+gr}^p$, 
where $r\equiv (k^2-K^2)/K^2$. After this we form $w_1$ by expanding
$w_N$ up to order $g$, and setting $\kappa=1$.
}
This leads to 
\begin{eqnarray}
w_1&=&\mu^2t^\gamma\Phi^2K^p\left[\bar f_0 \left(1-\frac{p}2
+ \frac{p}2\frac {1}{K^2}\right)+\bar f_1 \frac{y}{K^{q}}\right]\nonumber \\
&=&
\mu^ 2 t^\gamma\Phi^2 W_1(y).\label{@107c}
\label{@127}
\end{eqnarray}
The last term in the first line shows that for large $y$, $K$ has to grow like 
\begin{eqnarray}
K\propto y^{1/q}.
\label{@108a}\end{eqnarray}We now extremize $w_1$ with respect to $K$
and find that the derivative $\lfrac{dw_1}{dK}$ has to vanish, i.e.,
\begin{eqnarray}
\!\!\!\!\!\!
t^\gamma\Phi^2K^{p-1}\frac{p(2-p)}2
\left[\bar f_0\!
\left(1\!-\!\frac{1}{K^2}\right)-\bar f_1 c
\frac{y}{K^{q}}
\right]\!\!=\!0,
\label{@108}\label{@129}\end{eqnarray}
where 
\begin{eqnarray}
c\equiv\frac{2(p-q)}{p(p-2)}\approx 0.32.
\label{@121}\end{eqnarray}
In the free-particle limit $y\rightarrow 0$,
 the solution is $K(0)= 1$, and 
\begin{eqnarray}
w_1=\mu^2t^\gamma\Phi^2\bar f_0 .
\label{@129a}\end{eqnarray}

The last term in (\ref{@129}) shows once more 
that for large $y$, $K$ will be proportional to
$y^{1/q}$. 
Moreover, it allows to sharpen relation (\ref{@108a})
to the large-$y$ behavior
\begin{eqnarray}
K\rightarrow K_{\rm as}(y)=(cy)^{1/q}\approx 
0.648\, y^{0.381}.
\label{@123}\end{eqnarray}
Then the leading large-$y$ behavior 
of (\ref{@107}) is
$\Phi^2 (cy)^{p/q}\propto (\Phi^2) ^{p+1}$.

The first correction to the large-$y$ behavior comes from the second term in the brackets
of (\ref{@108}), which by 
(\ref{@123a})
should behave like
\begin{eqnarray}
K^2\rightarrow (c\,t^{-2\beta}g \Phi^2/\mu^2)^{\omega \nu/2\beta}
\approx
(c\,t^{-2\beta}g \Phi^2/\mu^2)^{0.76}
.\label{@111}\end{eqnarray}
Comparing this with (\ref{@110})
and (\ref{@108a}), we find 
$p=2(2-\eta)/\omega$ and $q=4\beta/\omega\nu$.
For the $N=2$ universality class these have 
the numerical values $p\approx 4.92$ and $q\approx
4\times 0.32/(0.8\times 0.66)\approx 2.63$. 
Solving
(\ref{@129}), we see that for small $y$, $K(y)$
has the diverging expansion
\begin{eqnarray}\!\!\!\!\!\!
K(y)\!=\!
1\!+\!0.32\, y\!-\!
0.165967\, y^2\!+\!0.155806 \,y^3
\!+\!\dots~,
\label{@}\end{eqnarray}
so that $W_1$
has the diverging expansion
\begin{eqnarray}
W_1=
1+ y+0.154436 y^3+\dots~.
\label{@}\end{eqnarray}
From Eq.~(\ref{@82}), 
we know that the power $p+1$ must be equal 
to $(\delta+1)/2$, so that
\begin{eqnarray}
y^{p/q}=y^{(\delta-1)/2}=y^{(2-\eta)\nu/2\beta}\approx 
y^{1.87}.
\label{@110B}\end{eqnarray} 

\begin{figure}[h]
\vspace{1.cm}
\hspace{-.9cm}
\unitlength.5pt
\hspace{-15em}
\begin{picture}(94.49,127.195)
\input epsf.sty
\def\dst{\displaystyle}
\def\IncludeEpsImg#1#2#3#4{\renewcommand{\epsfsize}[2]{#3##1}{\epsfbox{#4}}}


\put(220,0){\IncludeEpsImg{94.49mm}{64.13mm}{.30}{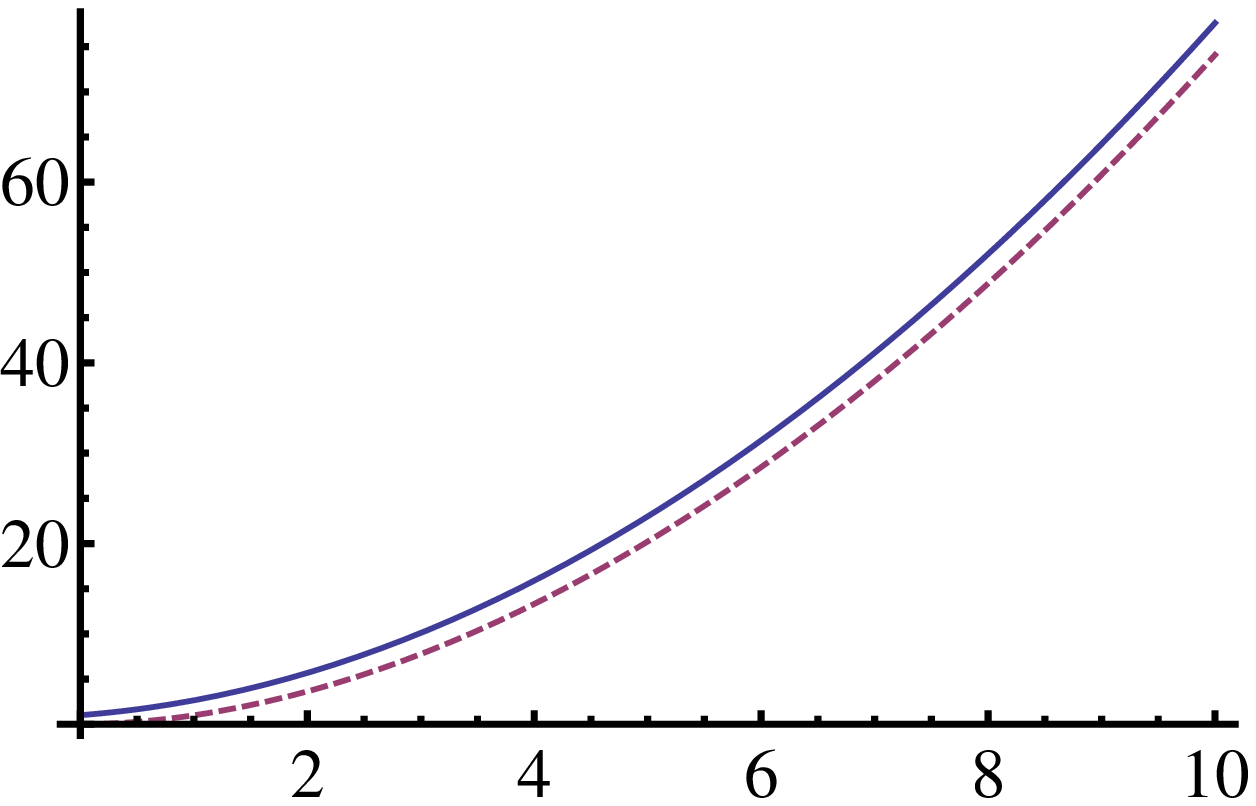}}
\put(-20,0){\IncludeEpsImg{94.49mm}{64.13mm}{.30}{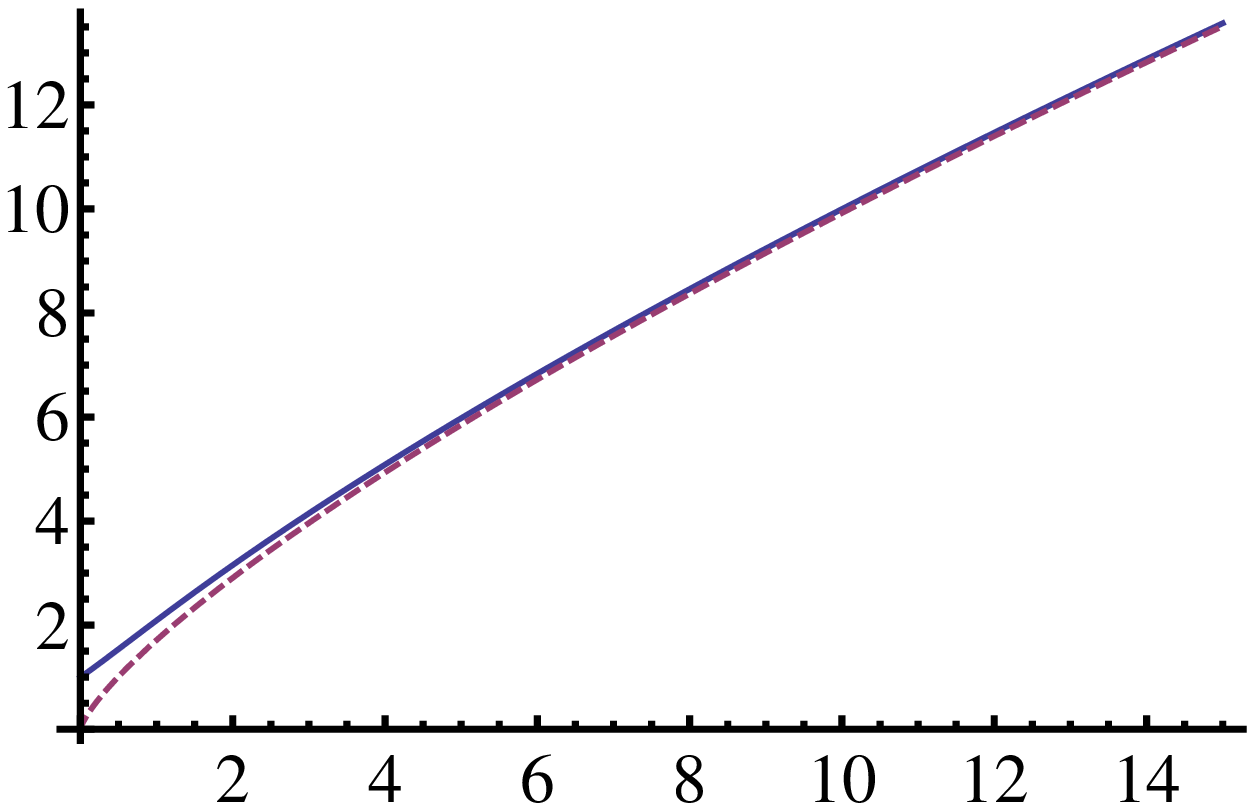}}
\put(78,-6){ $y$}
\put(318,-6){ $ y$}
\put(50,111){\fsz $K(y)$}
\put(285,131){\fsz $W_1(y)$}
\end{picture}
\caption[]{Solution of the variational equation (\ref{@108}) for $\bar f_1=1$.
The dotted curves show the pure large-$y$ behavior.
}
\label{KOFY}\end{figure}

If we insert (\ref{@108a})
into (\ref{@107}),  the extremal variational energy is 
\begin{eqnarray}
w_1(y)=\mu^2 t^ \gamma\Phi^2 K^p\left(1-\frac{p}2 + 
\frac{y
}{K^{q}}
\right).
\label{@133}\end{eqnarray}
where $K=K(y)$ is a function $y$ which is plotted in Fig.~\ref{KOFY}.

\comment{
\section{Effective Potential}
Let us now derive 
the effective potential (\ref{@WIDOM}).
To avoid an irrelevant infinite subtraction
in the total energy
we shall study only the  derivative $H_a(\Phi)=\partial v(\Phi)/\partial \Phi_a$
[recall (\ref{@HEQA})]
for which we find 
from (\ref{d12.109}) the expression
\begin{eqnarray}\!\!
H_a(\hspace{-1pt}\Phi\hspace{0pt})\hspace{-1pt}\!\!\!&=&\!\!\!
m^2 \hspace{-1pt}\!\left[\hspace{-1pt}1\!+\!
\frac {\lambda}{2}A_L(\hspace{-1pt}\Phi\hspace{-1pt})\!+\!
(N\!\!-\!1\hspace{-1pt}) \frac {\lambda}{6}A_T(\hspace{-1pt}\Phi\hspace{-1pt})
\!+\!\frac{g \Phi^2}{6m^2}\!\right]\!\Phi_a\nonumber \\
&+&\!\!\!
m^2 \hspace{-3pt}\left[\hspace{-1pt}
\frac{N\!+\!2}6\lambda\,
c_1
\!+\!\frac{N+8}{36}\lambda\frac{g \Phi^2}{m^2} c_2\right]\!\Phi_a,
\label{@EQOFS}\end{eqnarray}
where 
\begin{eqnarray}&&\!\!\!\!\!\!\!\!\!\!\!\!\!\!\!\!\!
A_L(\Phi)\!\!=\!\!{m^{\epsilon-2}}\!\!\int \dbar^D p\!\left[ \frac1{
p^2\!+\!m^2\!+\!(g/2)\Phi^2}\right.
\nonumber \\&&~~~~\left.~~~~~~~~~~~~~~~\!\!
- \frac1{
p^2\!+\!m^2}\!+\!\frac{g}2\Phi^2\frac1{
(p^2+m^2)^2}\right]\!,
\label{@}\end{eqnarray}
\begin{eqnarray}&&\!\!\!\!\!\!\!\!\!\!\!\!\!\!\!\!\!\!\!
A_T(\Phi)\!\!=\!\!{m^{\epsilon-2}}\!\!\!\int \dbar^D p\left[ \frac1{
p^2+m^2+(g/6)\Phi^2}\right.\nonumber \\&&~~~~\left.~~~~~~~~~~~~~
- \frac1{
p^2+m^2}\!+\!\frac g6\Phi^2\frac1{
(p^2+m^2)^2}\right]\!,
\label{@}\end{eqnarray}
and $c_1$, $c_2$ have been calculated in Eqs.~(\ref{d12.135a}), (\ref{@Incp'}),
(\ref{d12.135}) with their $1/\epsilon$-pole.
The first term renormalizes the mass to an expression 
of the form (\ref{@Form52}) with the critical exponent $\gamma$ 
 from (\ref{@Form91}). The second divergence is used to 
renormalize the coupling constant
as in (\ref{@Form89}).
\comment{
\begin{eqnarray}
c_1=m^{\epsilon-2}\int \db ^Dp\frac{1}{p^2+m^2},\nonumber \\
c_2=m^{\epsilon}\int \db ^Dp\frac{1}{(p^2 +m^2)^2},
\label{@}\end{eqnarray} 
Thus the effective action reads
\begin{eqnarray}
&&\!\!\!\!\!\!\!\!\!\!\!\!\!\!
{\cal A}^{\rm eff}=\frac{m^2}2\Phi^2+\frac{g}{4!}\Phi ^4+\frac1 2\int \frac{d^Dp}{(2\pi)^D}\bigg[
(p^2+m^2)\delta_{ij}+\frac{g}{6}(\delta_{ij}\Phi^2+2 \Phi_i\Phi_j)\bigg].
\label{@}\end{eqnarray}
}
Introducing the variable
 $y\equiv g \Phi^2/m^2$, these become
\begin{eqnarray}&&\!\!\!\!\!\!\!\!\!\!\!\!\!\!\!\!\!\!\!
A_L(y)\!\!=\left(1+\frac{y}2\right)\log \left(1+\frac{y}2\right)-
\frac{y}2,
\label{@}\end{eqnarray}
and
\begin{eqnarray}&&\!\!\!\!\!\!\!\!\!\!\!\!\!\!\!\!\!\!\!
A_T(y)\!\!=\!\!
=\left(1+\frac{y}6\right)\log \left(1+\frac{y}6\right)-
\frac{y}6.
\label{@}\end{eqnarray}
The
the effective potential whose derivative is
(\ref{@EQOFS}) reads
\begin{eqnarray}
\!\!\!\!\!\!\!\!
\!\!\!&&v(\Phi)=m^4g{}^{-1}\Bigg\{\frac{y}2+\frac{y^2}{24}
-\frac{3N}{2(4\!-\!N)}
\!\!\!\!\!\!\!\!\!
\nonumber\\
&&~~~~~~~~~~~~
+
\frac{g}{16}(N\!-\!1)\left(
12+\frac y 6\right)\bigg[\!1\!+2\log\left(1\!+\!\frac y6\right)^2\bigg]
\nonumber\\
&&~~~~~~~~~~~~~~\!\!\!\!+\frac{g}{16}\left(
1+\frac y 2\right)^2\bigg[1\!+\!2\log\left(1\!+\!\frac y2\right)\bigg]\Bigg\}.
\label{@15.68x}\end{eqnarray}
Going here to the critical limit $y\rightarrow \infty$,
we can recover power behavior (\ref{@Form79}). 
}

\comment{
It is instructive 
to determine from this the
complete phenomenological scaling properties
of the system.
First we deduce from 
$v(\Phi)$ the equation of state
a la Widom by forming the derivative 
\begin{eqnarray}
H\equiv  -\partial v(\tilde \Phi)/\partial \tilde \Phi.
\label{@}\end{eqnarray}
Introducing 
the variable
\begin{eqnarray}
x\equiv m^2/y^{1/2\beta},~~~
\label{@}\end{eqnarray}
we find
\begin{eqnarray}\!\!\!\!\!\!\!\!\!\!\!\!\!\!\!\!
H\!\!\!&\propto&\!\!\!
\Phi^\delta\Bigg\{
(x+\sfrac16)
+\frac{3\epsilon}{N+8}
\Bigg[\frac{N-1}6
 \\&&\!\!\!\!\!\!\!\times(x+\sfrac1 6)[\log(x+\sfrac1 6)\!+\!1]
\!+\!\frac{1}2(x+\sfrac1 2)[\log(x+\sfrac1 2)\!+\!1]
\Bigg]\!\Bigg\}.\nonumber 
\label{@15.68x}
\end{eqnarray}
where the variable $x$ is related to $y$ by the scaling relation
\begin{eqnarray}
x\equiv m^2/y^{1/2\beta}.
\label{@SCR}\end{eqnarray}
\comment{According to Widom, this may be written as
\begin{eqnarray}
H=f(x),~~~x\equiv m^2/y^{1/2\beta}.
\label{@}\end{eqnarray}
}
To first order in $\epsilon$,
the relation is
\begin{eqnarray}
x=\frac{m^2}y\left[ 1-\left(\frac1{2\beta}-1\right)\log y+\dots\right],
\label{@}\end{eqnarray}
or
\begin{eqnarray}
y=\frac{m^2}y\left[ 1+\left(\frac1{2\beta}-1\right)\log x+\dots\right],
\label{@}\end{eqnarray}
where
\begin{eqnarray}
\frac1{2\beta}-1=\frac{3\epsilon}{N+8}.
\label{@}\end{eqnarray}
Inserting this into 
(\ref{@15.68x}) we find
\begin{eqnarray}
H=\tilde\Phi^\delta f(x),
\label{@}\end{eqnarray} 
with
\begin{eqnarray}
\delta=3+ \epsilon,
\label{@}\end{eqnarray}
and
\comment{
\begin{eqnarray}
H=M^\delta f(x),
\label{@}\end{eqnarray} 
where 
}
\begin{eqnarray}
f(x)=1+x+\epsilon f_1(x),
\label{@}\end{eqnarray}
where 
\begin{eqnarray}
&&\!\!\!\!\!\!\!\!\!\!\!\!\!\!\!\!\!\! f_1(x)=\frac{1}{2(N+8)}[(N-1)(x+1)\log (x+1)\nonumber 
\\&&\!\!\!\!\!\!+3(x+3)\log (x+3)-9(x+1)\log 3 
+6 x \log 2].
\label{@}\end{eqnarray}
}

\comment{
The kinetic term $ (\hat{\bf p}^2)^{1-\eta/2}$ in (\ref{@GPP})
[and of course (\ref{@fGP})]
is modified to
$ (\hat{\bf p}^2)^{1-\eta/2}[1+{\rm const}\times\xi^{-\omega
}(\hat{\bf p}^2)^ {-\omega/2}+\dots]$ \cite{REMDEL}, 
 the
interaction term $|\Psi|^{\delta-1}$ to $|\Psi|^{\delta-1}(1+{\rm const}\times 
\xi^{-\omega}|\Psi|^{-\omega \nu/\beta})$.
}

For large $y$, where $K(y)$ has the limiting behavior (\ref{@123}),
$W_1$ becomes
\begin{eqnarray}
W_1(y)&\rightarrow&W_{\rm as}(y)=K^p\left(1-\frac{p}2 +\frac1c \right)
\nonumber \\
&\approx& 0.197\, y^{1.87}. 
\label{@133B}\end{eqnarray}
Near the limit, the  corrections to (\ref{@123}) are
\begin{eqnarray}
K(y)&=&K_{\rm as}(y)\left(1+\sum_{m=1}^\infty\frac{h_m}{y^{m\omega\nu/2\beta}} \right)
\nonumber \\&
\approx&
0.648\, y^{0.381}\left(1+\sum_{m=1}^\infty \frac{h_m}{y^{ 0.761\,m}} \right).
\label{@}\end{eqnarray}
with 
$h_1\approx0.909$,\,
$h_2\approx-0.155, \dots~.$
Inserting this into (\ref{@127}), we find
\begin{eqnarray}
W_1(y)&=&W_{\rm as}(y)\left(1+\sum_{m=1}^\infty\frac{ b_m}{y^{m\omega\nu/2\beta}} \right)
\nonumber \\&
\approx&
0.197\, y^{1.87}\left(1+\sum_{m=1}^\infty \frac{b_m}{y^{ 0.761\,m}} \right).
\label{@}\end{eqnarray}
with 
$b_1\approx3.510,~b_2\approx4.65248,\dots~.$

\section{Fractional Gross-Pitaevskii Equation}

We now extremize the effective action 
(\ref{d12.127}) with the two-loop corrected quadratic term
(\ref{d12.130}). We consider the case of $N=2$ where $\Phi^2=\Psi^*\Psi$.
Then we take the effective potential 
(\ref{@133})
with the extremal 
 $K=K(y)$ as a function $y$ plotted in Fig.~\ref{KOFY}.
From this we form the derivative 
$\partial w_1(y)/\partial \Psi^*$
and obtain the 
time-independent
{\it fractional
Gross-Pitaevskii equation\/}:
\begin{eqnarray}\!\!\!\!\!
(\hat{\bf p}^2) ^{1-\eta/2}\Psi\! +
\frac{\partial w_1(y)}{\partial \Psi^*}
\!=\!0.
\label{@fGP}\end{eqnarray}
If we use the weak-coupling limit
of $w_1(y)$ and the gradient term, this reduces to the ordinary 
time-independent
Gross-Pitaevskii equation
\begin{eqnarray}
\left[
-\frac{\hbar ^2}{2M}\nablab^2-\mu+g_S\Psi^\dagger\Psi\right]\Psi(x)=0.
\label{@GPEO}
\label{@22}\end{eqnarray}
In a harmonic trap
$\mu$ is replaced by
$\mu+M \omega^2 {\bf x}^2/2$.
Recalling the relation 
$m^2=-2M\mu$
one has 
$m^2=-2M\,M\omega^2R_c^2(1-R^2/R_c^2)$.
The oscillator energy 
$M\omega^2R_c^2$ corresponds to a length scale $\ell_{\cal O}$
by the relation $M\omega^2R_c^2=\hbar  ^2/M\ell_{\cal O}^2$,
so that $m^2$ may be written as 
$m^2=\mu^2(R^2/R_c^2-1)$ with $\mu=1/\ell_{\cal O}$.

In the strong-coupling limit, however,
we arrive
at the
time-independent
{\it fractional
Gross-Pitaevskii equation\/}:
\begin{eqnarray}\!\!\!\!\!
 \left[
(\hat{\bf p}^2) ^{1-\eta/2}\! +\frac{\delta\!+\!1}{4 \mu^\eta}
g_c
|\Psi({\bf x})|^{\delta-1}\!\right]\!\!\Psi({\bf x})\!=\!0.
\label{@fGPE}\end{eqnarray}
By using the full effective action for all coupling strengths 
and masses $m^2$ we can calculate the properties of the condensate at an coupling strength.
Before reaching  the strong-coupling limit,
we may use 
Eq.~(\ref{@fGP}) to calculate the field strength
as a function of $g_S$.
Then the critical exponents $\eta, \alpha,\beta,\delta$ have not yet reached 
their strong-coupling values 
by must be replaced by the $\bar g_B$-dependent 
precritical values calculated from
(\ref{@grenofgbar1})
and the corresponding equations 
for $\eta, \alpha,\beta,\delta$.

In a trap, the mass term becomes weakly space-dependent.
If the trap is rotationally symmetric, then $m^2$
will depend on $R=|{\bf x}| $
and the time-independent Gross-Pitaevskii equation
has to be solved with $m^2(R)\propto 1-R^2/R_c^2$.
More specifically,
the bare coupling constant 
on the right-hand side has to be 
determined 
in such a way that
$m^2/m_B^2$ has the experimental size.
If the experiments are performed in an external magnetic field $B$
the $s$-wave scattering length
$a_s$ has an enhancement factor 
$(B/B_c-1)^{-1}$
and Eq.~(\ref{@grenofgbar1})
can again be used.
We can then calculate the density profile
quite easily in the Thomas-Fermi approximation 
as done in Ref.~\cite{BSE}.

\begin{figure}[h]
\vspace{1.cm}
\hspace{-.6cm}
\unitlength.5pt
\hspace{-15em}
\begin{picture}(94.49,67.195)
\input epsf.sty
\def\dst{\displaystyle}
\def\IncludeEpsImg#1#2#3#4{\renewcommand{\epsfsize}[2]{#3##1}{\epsfbox{#4}}}
\put(-20,0){\IncludeEpsImg{94.49mm}{64.13mm}{0.300}{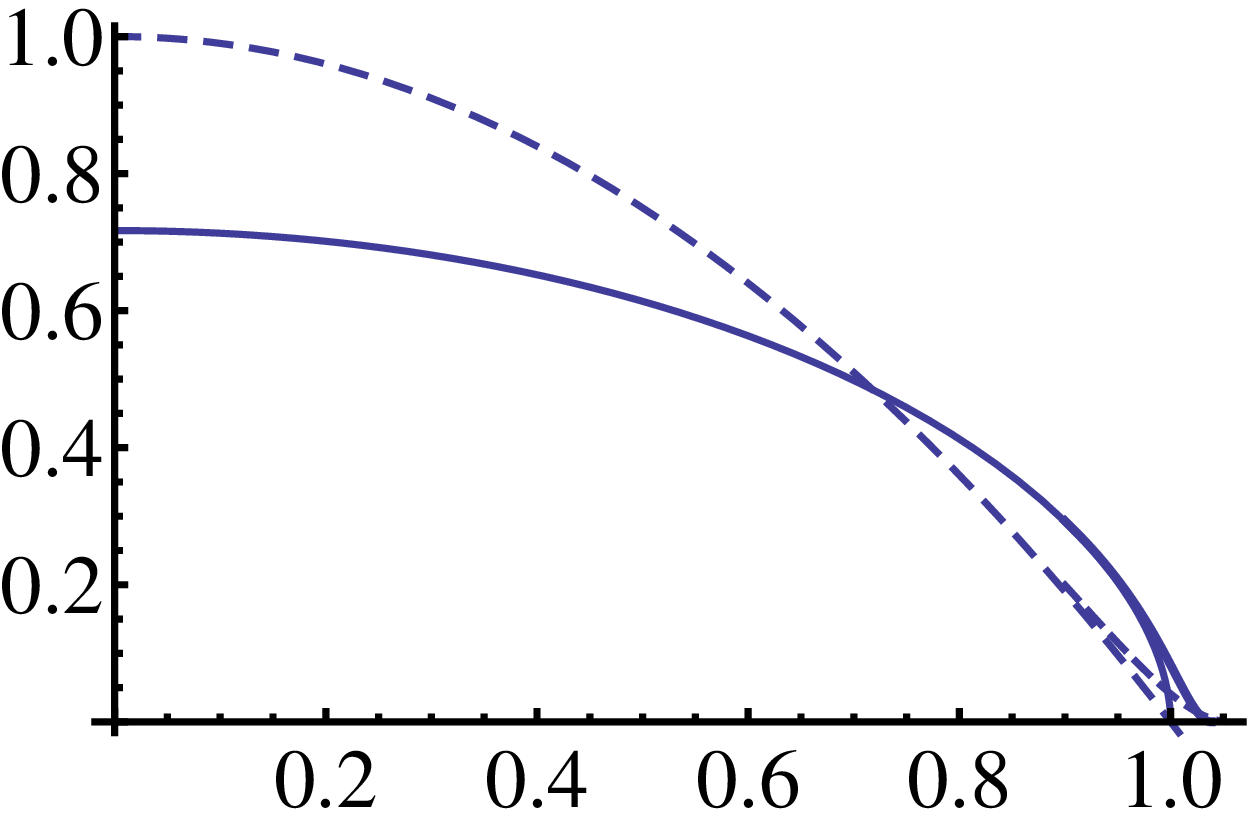}}
\put(220,0){\IncludeEpsImg{94.49mm}{64.13mm}{0.300}{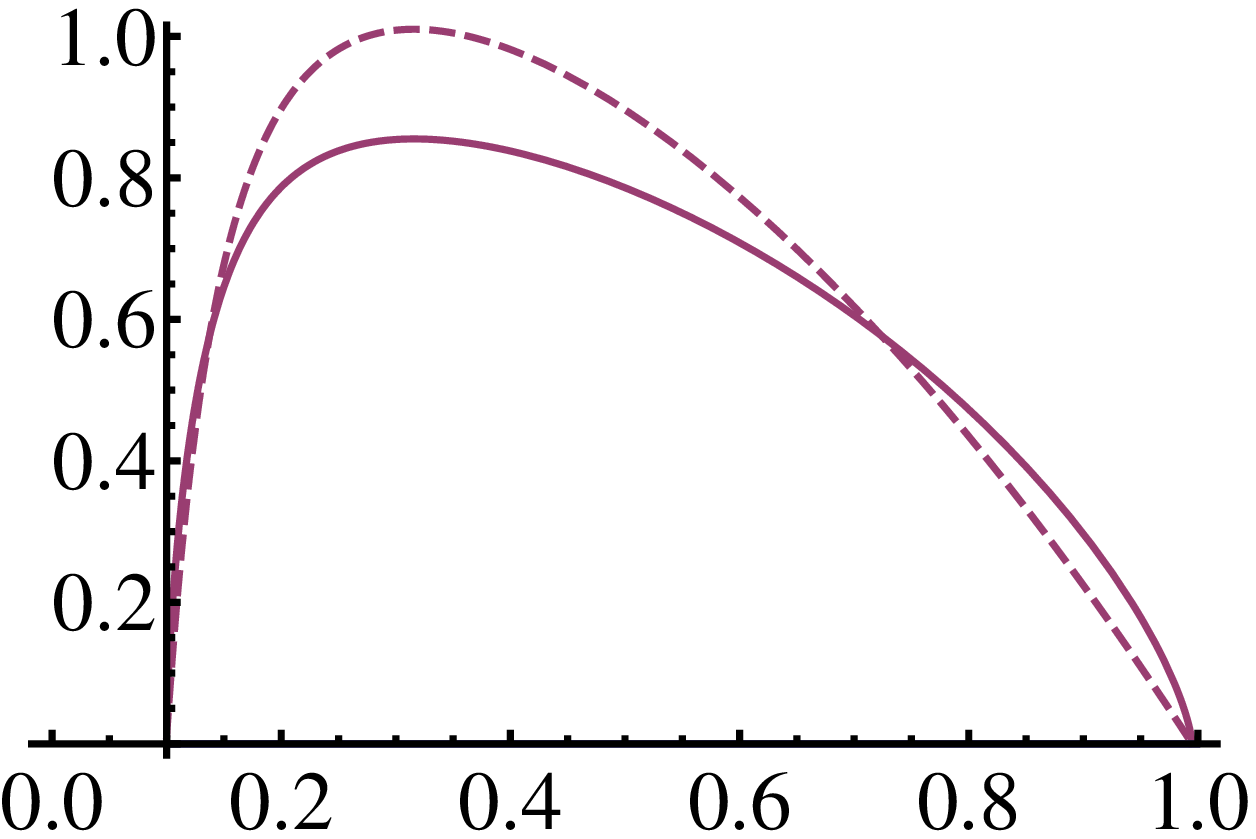}}
\put(78,-6){ $ r$}
\put(318,-6){ $ r$}
\put(50,111){\fsz GP}
\put(50,80){\fsz FGP}
\put(285,131){\fsz GP}
\put(285,105){\fsz FGP}
\put(48,40){\fsz $\rho=\Psi^*\Psi$}
\put(288,50){\fsz $\rho=\Psi^*\Psi$}
\end{picture}
\caption[]{Condensate density
from Gross-Pitaevskii equation
(\ref{@GPE}) (GP,dashed) and its fractional version 
(\ref{@fGPE} (FGP), both in Thomas-Fermi approximation
where the gradients are ignored. 
The FGP-curve shows a marked 
depletion of the condensate.
On the right hand, a vortex is included.
The zeros at $r\approx 1$ will be smoothened by
the gradient terms in (\ref{@GPE}) and 
(\ref{@fGPE}), as shown 
on the left-hand plots without a vortex. 
The curves can be compared with those in Ref.~\cite{DEV1,FD,HAU,DEV2,DEV3}. 
}
\label{@f0}\end{figure}

In a rotating 
BEC we can calculate the different forms of the 
density profiles of vortices 
for various coupling strengths which can be varied 
from weak to strong by subjecting the
BEC to different magnetic fields, raising it from zero
up to the Feshbach resonance.
The profiles are shown in Fig. \ref{@f0}).

In addition, the central region is depleted (see Fig.~\ref{@f0}).

\section{Summary}

We have shown that the expansion of the effective action 
of a $\phi^4$-theory
in even powers of the field strength $\Phi=\langle \phi\rangle$
can be be resummed
to obtain an expression that is valid for {\it any} field strength, even 
in the 
the strong-coupling limit.
It has the phenomenological scaling form once proposed by
Widom, and can be used to calculate the shape of a BEC
up to the Feshbach resonance, 
with and without rotation.


\begin{thebibliography}{9}
\bibitem{remn}
A travelling  pedestrian
salesman is a Gaussian
random walker, as a jetsetter he becomes a L\'evy random walker.
\bibitem{METZ}
The relevant functional matrix is
$\langle{\bf x} |
(-\nablabf^2)^{\lambda/2}|{\bf x} '\rangle=\Gamma[-\lambda/2]^{-1}
\int d\sigma\, \sigma^{-\lambda/2-1}
{(4\pi \sigma)^{-D/2}}e^{R^2
/4\sigma}
$=
$\!\!{}^D\! c_\lambda R^{-\lambda-D}$, where
${}^D\!c_\lambda=2^{\lambda}\Gamma((D+\lambda)/2)/\pi^{D/2}\Gamma(-\lambda/2),$
and $R\equiv|{\bf x}-{\bf x}'|$. If $\lambda$ is close to an
even integer, it needs a small positive shift $\lambda\rightarrow\lambda_+\equiv\lambda+\epsilon$ and we can replace $ \epsilon R^{\epsilon-1}/2 $ by 
$\delta(R)=S_DR^{D-1}\delta^{(D)}({\bf R})$.
For  $A>0$
 we have $|{\bf x}'
|^{-A}= {}^D\!c_{\lambda_A}^{-1}\langle{\bf x}' |
(-\nablabf^ 2)^{\lambda_A/2}|{\bf 0} \rangle$ with
$\lambda_A\equiv A-D$, 
so that we find
$\int d^Dx'\langle{\bf x} |(-\nablabf^ 2)^{\lambda/2}|{\bf x}' \rangle |{\bf x}'|^{-A}=
{}^D\!c_{\lambda_A}^{-1}\langle{\bf x}|(-\nablabf^ 2)^{(\lambda+A-D)/2}
|{\bf 0}\rangle =
{}^D\!c_{\lambda+A-D}{}^D\!c^{-1}_{\lambda_A}|{\bf x}
|^{-A-\lambda}$.
\bibitem{REM1}
This technique is explained in Chapters~12 and 19 of Ref.~\cite{PI}.
The pseudotime $s$ resembles the so-called
 Schwinger proper time used in relativistic physics.
\bibitem{REMAN}
There should be  no danger of confusing the fluctuating
noise variable $\eta$
in this equation  with the constant
critical exponent $\eta$ in (\ref{d12.130P}).
\bibitem{REMAR}
The decimal numbers are from seven-loop calculation in $D=3$ dimensions 
in Table 20.2 of Ref.~\cite{KS22}.
\end{thebibliography}
\end{document}